%% file: main.tex
\documentclass[final,5p,times,twocolumn]{elsarticle}

\usepackage[utf8]{inputenc}
\input{definitions.tex}
\usepackage{graphicx}
\usepackage{multicol}
\usepackage{amsmath}
\usepackage{amssymb}
\usepackage{color}
\usepackage{epstopdf}
\usepackage{bm}
\usepackage{mathtools}
\usepackage{array}
\usepackage{accents}
\usepackage{adjustbox}
\usepackage{lscape}
\usepackage{floatpag}
\usepackage{subcaption}
\usepackage{siunitx}
\usepackage[ruled,vlined,linesnumbered]{algorithm2e}
\usepackage{multirow}
\usepackage{makecell}
\usepackage{placeins}
\usepackage{lineno,hyperref}
\usepackage{adjustbox}
\usepackage{rotating}
\modulolinenumbers[5]

\DeclareMathOperator{\IAE}{IAE}
\DeclareMathOperator{\IAC}{IAC}
\DeclareMathOperator{\IACD}{IACD}
\DeclareMathOperator{\IADEE}{IADEE}

\newtheorem{remark}{Remark}
\newtheorem{assumption}{Assumption}
\newtheorem{theorem}{Theorem}

\journal{ISA Transactions}

\begin{document}

\begin{frontmatter}



\title{Tuning of extended state observer with neural network-based control performance assessment}


\author[PUT]{Piotr Kicki\corref{mycorrespondingauthor}}
\ead{piotr.kicki@put.poznan.pl}

\author[PUT]{Krzysztof {\L}akomy\corref{mycorrespondingauthor}}
\cortext[mycorrespondingauthor]{Corresponding author}
\ead{krzysztof.lakomy92@gmail.com}

\author[UTS]{Ki Myung Brian Lee\color{black}}
\ead{brian.lee@student.uts.edu.au}

\address[PUT]{Poznań University of Technology, Piotrowo 3A, 60-965, Poznań, Poland}
\address[UTS]{University of Technology Sydney, 15 Broadway, Ultimo NSW 2007, Sydney, Australia}

\begin{abstract}
The extended state observer (ESO) is an inherent element of robust observer-based control systems that allows estimating the impact of disturbance on system dynamics. Proper tuning of ESO parameters is necessary to ensure a good quality of estimated quantities and impacts the overall performance of the robust control structure. In this paper, we propose a neural network (NN) based tuning procedure that allows the user to prioritize between selected quality criteria such as the control and observation errors and the specified features of the control signal. The designed NN provides an accurate assessment of the control system performance and returns a set of ESO parameters that delivers a near-optimal solution to the user-defined cost function. The proposed tuning procedure, using an estimated state from the single closed-loop experiment produces near-optimal ESO gains within seconds.

\end{abstract}




\end{frontmatter}



\section{Introduction}

The extended state observer (ESO) is an inherent component of the robust control framework that relies on the cancellation of disturbances using their lumped estimate in the feedforward component of the robust control law. The general idea of such control structure was utilized in many specific robust algorithms such as active disturbance rejection control (ADRC) \cite{han2009,gao2006}, disturbance observer based control (DOBC) \cite{chen2016,li2014}, or robust observer based control \cite{isidori2017}, while its applicability has been proven in many fields including power electronics \cite{lakomy2021,madonski2019,yang2018}, temperature control \cite{zheng2018}, motion control \cite{ramirez2014,xue2017}, and robotics \cite{neria2014,lakomy2021ejc}. Besides the fact that there is a wide variety of ESO architectures that deal with disadvantages of a most commonly used Luenberger-like extended high-gain observer (HGO) \cite{zheng2012, freidovich2006} in terms of the general disturbance observation quality \cite{sun2016_2}, transient performance \cite{wei2019}, or the robustness to measurement noise \cite{wang2017,lakomy2021isat}, the final characteristics of the control system performance depend greatly on the appropriate tuning of particular observer parameters. 

The most common parametrization of HGOs utilized within the robust control scheme is the so-called bandwidth parametrization \cite{gao2006scaling} that is based on the choice of a single parameter value interpreted as an eigenvalue of the designed closed-loop observation subsystem. 
Despite its indisputable advantages, like intuitiveness and easy implementation, tuning with bandwidth parameterization may lead to suboptimal performance, in terms of control errors, energy consumption, and the impact of measurement noise, as only one degree of freedom is available. 
In the literature, many methods have been proposed for tuning bandwidth-parameterized observers.
In \cite{zhang2019} and \cite{wang2021}, the authors presented an analytical tuning method providing the best performance of the ADRC structure expressed solely upon the control-error-dependent criteria in a noiseless environment. In \cite{wang2019} and \cite{deboon2021}, the authors considered also the control cost as a factor that needs to be minimized to reduce the energy consumption of the robust control process, while in \cite{madonski2013} the observation error of the measured signals was taken into account. Tuning procedures described in \cite{nowak2020} and \cite{grelewicz2020} have utilized prior knowledge about the plant structure and some known or identified model parameters to obtain assumed control performance requirements. In \cite{nowak2020} and \cite{herbst2020}, the authors presented an observer tuning method that is relative to gains of the selected ADRC controller. Some methods consider automatic tools designed for tuning the overall ADRC structure, including observer gains, to satisfy some predefined criteria determining the robustness of the control structure \cite{sun2016}. Whereas \cite{xue2016} describes the constraints that should be satisfied by the observer gains to achieve stability in a discrete-time ADRC control structure.


For a few decades neural networks have been utilized along with the classical control solutions as a part of the overall control system \cite{nn_control_survey}. In \cite{nn_speed_estimation} and \cite{gaining_sense_of_touch} authors used neural networks to estimate quantities that are hard to measure directly from other readily available measurements. The same idea is the essence of the neural network-based observers \cite{nn_observer_1, nn_observer_2}, which adaptively track signals that are crucial for controlling the system. Neural networks, which are proven to be universal approximators, are also used to approximate the dynamics of the systems that are difficult to model or parts thereof \cite{nn_model_1, nn_model_2}, allowing the use of the well established model-based methods. Furthermore, neural networks can serve as a standalone adaptive controller \cite{nn_control_1, nn_control_2}, however, they usually come with no warranties about their control performance or robustness to noise and external disturbances, limiting their use in safety-critical applications. Instead of relying on a standalone neural network based controller, neural networks can be used to tune some well known control schemes, like PID \cite{nn_tuning_pid} or ADRC \cite{nn_tuning_adrc_1, nn_tuning_adrc_2, nn_tuning_adrc_3}. This allows taking advantage of their robustness while improving the control quality by a data-driven choice of the controller or observer gains.
Authors of \cite{nn_tuning_adrc_2} proposed a reinforcement learning based approach to tune the gains in the nonlinear state error feedback control law.
Whereas in \cite{nn_tuning_adrc_1} a radial basis function neural network was used to estimate the ESO gains in an online manner to minimize the square of the difference between the actual and desired output of the system. A similar application of neural networks was presented in \cite{nn_tuning_adrc_3}, where a fully connected neural network was used. However,
both \cite{nn_tuning_adrc_1} and \cite{nn_tuning_adrc_3} did not consider measurement noise, what hinders appropriate choice of observer gains.

Most of the aforementioned tuning methods minimize some predefined cost function that determines which performance criteria should be optimized. In this work, we propose a different approach to tune the extended state observer using a neural network. Instead of training a neural network to choose gains directly, our method uses a neural network as a performance estimator of the closed-loop system.
We train a neural network performance estimator on a dataset of experiments, to estimate values of several performance criteria, which take into account not only a tracking quality but also a control effort and measurement noise suppression.
The proposed performance assessment network, estimates the expected performance of the system given the specific observer gains, within some used-defined regions of the state, control signals, and initial conditions. Using this estimator for multiple different proposed sets of the ESO gains that are not limited to the previously described bandwidth-parametrization, it is possible to choose
the one which minimizes chosen criterion and assures near-optimal performance. 
The introduced method allows quickly tuning the ESO gains according to a user-defined criterion based on a single closed-loop experiment, without the need for multiple system simulations nor open-loop model identification.

The main contribution of this paper is threefold:
\begin{enumerate}
    \item a novel application of neural networks to estimate the multi-criteria performance of the control system,
    \item a new ESO tuning method that selects a set of near-optimal gains using the estimates of the expected performance, without the constraints of bandwidth parameterization,
    \item an intense numerical verification of the proposed method, its generalization abilities, as well as the impact of the criterion formula on the resultant behavior of the considered systems.
\end{enumerate}

In the course of our analysis, we showed that bandwidth parametrization captures the entire spectrum of the performance for the considered systems and performance criteria. This implies that bandwidth parametrization allows choosing near-optimal gains for the user-defined performance criterion, however one has to carefully tune the bandwidth value. Moreover, we developed a publicly available dataset\footnote{\url{https://drive.google.com/file/d/1I2iVfog2ovytaaElmCYD6arFoj_8weHy/view}} of closed-loop systems simulations for two different control object structures with changing parameters, initial conditions, noise standard deviation, and ESO gains.




\textbf{Notation:} Throughout this paper, we use $\realNumbers$ as a set of real numbers, $\nonNegativeRealNumbers$ as a set of non-negative real numbers, $\negativeRealNumbers$ as a set of negative real numbers, $\positiveIntegerNumbers$ as a set of positive integers, and $\complexNumbers$ as a set of complex numbers. A ball $\ballDomain{r}$ of radius $r>0$ is defined as $\ballDomain{r}\triangleq\{\stateVector\in\realNumbers^n: \|\stateVector\|\leq r\}$ for $n\in\positiveIntegerNumbers$, while set $\continuousFunctionSet{1}$ represents a class of locally Lipschitz continuously differentiable functions. Zero and identity matrices are, respectively, represented as $\zeroMatrix\in\realNumbers^{m\times n}$ and  $\identityMatrix\in\realNumbers^{n\times n}$ for $m>0$ and $n>0$, while $\stateMatrix{n} \triangleq \begin{bmatrix} \zeroMatrix^{(n-1) \times 1} & \identityMatrix^{(n-1)\times (n-1)} \\
0 & \zeroMatrix^{1\times (n-1)}\end{bmatrix}\in\realNumbers^{n\times n}$, $\inputVector{n} \triangleq \left[\zeroMatrix^{\color{black}1\times (n-1)\color{black}} \ 1\right]^\top\in\realNumbers^n$,  $\outputVector{n} \triangleq \left[1 \ \zeroMatrix^{\color{black}1\times  (n-1)\color{black}}\right]^\top\in\realNumbers^n$, and $\inputVectorExtendedState{n+1} \triangleq \left[\zeroMatrix^{\color{black}1\times (n-1)\color{black}} \ 1 \ 0\right]^\top\in\realNumbers^{n+1}$. Symbol $\eigenvalue{i}(\pmb{A})$ represents the $i$-th eigenvalue of matrix $\pmb{A}$, $\realPart{x}$ is the real part of a complex number $x\in\complexNumbers$, $\pmb{A}\succ0$ means that matrix $\pmb{A}\in\realNumbers^{n\times n}$  is positive definite in the sense that $\pmb{x}^\top\pmb{A}\pmb{x}>0$ for all $\pmb{x}\in\realNumbers^n$. Throughout the paper, accent $\hat{x}$ corresponds to the estimated value of signal $x$, while accent $\tilde{x}$ determines the observation error of signal $x$, i.e., $\tilde{x}\triangleq x-\hat{x}$.

\section{Preliminaries}
\label{sec:prelims}

\subsection{Control plant description and conventional robust control design}
Let us consider a second-order nonlinear model of the dynamical system, described with a general equation 
\begin{align}
    \begin{cases}
        \stateVectorDerivative(t) = \stateMatrix{2}\stateVector(t) + \inputVector{2}\left[\wenchaoDriftVector(\stateVector,t) + \wenchaoInputVectorField(\stateVector,t)\controlSignal(t) +  \externalDisturbance(t)\right] \\ 
        \systemOutput(t) = \outputVector{2}^\top\stateVector(t) + \measurementNoise(t),
    \end{cases}
    \label{eq:originalSystem}
\end{align}
where $\stateVector\triangleq[\stateVectorElement{1} \ \stateVectorElement{2}]^\top\in\realNumbers^2$ is the state vector, $\wenchaoDriftVector(\stateVector,t)$ is the drift vector field describing system dynamics, $\wenchaoInputVectorField(\stateVector,t)$ is the vector field representing the input gain, $\controlSignal(t)\in\realNumbers$ is the control signal, $\externalDisturbance(t)\in\realNumbers$ is a bounded external disturbance affecting the system, $\systemOutput(t)\in\realNumbers$ is the system output, and $\measurementNoise(t)\in\realNumbers$ is the measurement noise.

\begin{assumption}
    \label{ass:1}
    The state domain of system \eqref{eq:originalSystem} is bounded to an arbitrarily large compact domain, such that $\stateVector\in\ballDomain{\stateBallRadius}$, for $0<\stateBallRadius<\infty$.
\end{assumption}

\begin{assumption}
    External disturbance $\externalDisturbance(t)\in\continuousFunctionSet{1}$ and its derivative belong to the bounded domains $\externalDisturbance(t)\in\ballDomain{\externalDisturbanceBallRadius}$ and $\externalDisturbanceDerivative\in\ballDomain{\externalDisturbanceDerivativeBallRadius}$ for $0<\externalDisturbanceBallRadius,\externalDisturbanceDerivativeBallRadius<\infty$.
\end{assumption}

\begin{assumption}
    \label{ass:noiseBoundedness}
    The measurement noise has zero-mean and is bounded, i.e., $\measurementNoise\in\ballDomain{\measurementNoiseBallRadius}$ for $0<\measurementNoiseBallRadius<\infty$. 
\end{assumption}

\begin{assumption}
    Drift vector field $\wenchaoDriftVector(\stateVector,t):\ballDomain{\stateBallRadius}\times\nonNegativeRealNumbers\rightarrow\ballDomain{\driftVectorBallRadius}$ and input gain vector field $\wenchaoInputVectorField(\stateVector,t):\ballDomain{\stateBallRadius}\times\nonNegativeRealNumbers\rightarrow\ballDomain{\inputGainBallRadius}$ are locally Lipschitz continuous functions, i.e. $\wenchaoDriftVector,\wenchaoInputVectorField\in\continuousFunctionSet{1}$, for $0<\driftVectorBallRadius,\inputGainBallRadius<\infty$.
\end{assumption}

\begin{remark}
    In this paper, we will be modelling the measurement noise with a zero-mean truncated normal distribution $\measurementNoise\sim\normalDistribution(0,\measurementNoiseStandardDeviation^2)$, where $\measurementNoiseStandardDeviation$ is a measurable standard deviation, while the truncation borders are determined by $\measurementNoiseBallRadius$ introduced in Assumption \ref{ass:noiseBoundedness}. 
\end{remark}

System \eqref{eq:originalSystem} can be rewritten as:
\begin{align}
    \begin{cases}
        \stateVectorDerivative(t) = \stateMatrix{2}\stateVector(t) + \inputVector{2}[\wenchaoInputVectorFieldEstimate\controlSignal(t) + \underbrace{\wenchaoDriftVector(\stateVector,t) + [\wenchaoInputVectorField(\stateVector,t)-\wenchaoInputVectorFieldEstimate]\controlSignal(t) +  \externalDisturbance(t)}_{\totalDisturbance(\stateVector,t)}] \\ 
        \systemOutput(t) = \outputVector{2}^\top\stateVector(t) + \measurementNoise(t),
    \end{cases}
    \label{eq:rewrittenOriginalSystem}
\end{align}
where $\wenchaoInputVectorFieldEstimate$ is a constant approximation of the input gain $\wenchaoInputVectorField(\stateVector,t)$, while $\totalDisturbance(\stateVector,t)$ represents the so-called total disturbance that lumps external disturbances, unmodeled dynamics, and the impact of the parametric uncertainty of the input gain. 

\begin{assumption}
    \label{ass:5}
    Following the results presented in \cite{xue2018} and \cite{chen2020}, the
    input parameter estimate $\wenchaoInputVectorFieldEstimate$ is a rough approximation of $\wenchaoInputVectorField(\stateVector,t)$, satisfying 
    \begin{align}
        \forall_{t\geq0}\frac{\wenchaoInputVectorField(\stateVector,t)}{\wenchaoInputVectorFieldEstimate}\in\left(0,2+\frac{2}{n}\right),
        \label{eq:inputGeinRelation}
    \end{align}
    where $n=2$ is the order of system \eqref{eq:originalSystem}. 
\end{assumption}

To adjust the representation of system dynamics into a form allowing the design of a disturbance observer based control structure, let us introduce an extended state 
\begin{align}
    \label{eq:extendedStateVector}
    \extendedStateVector=[\extendedStateVectorElement{1} \ \extendedStateVectorElement{2} \ \extendedStateVectorElement{3}]^\top\triangleq[\stateVector^\top \ \totalDisturbance]^\top\in\realNumbers^3,
\end{align}
which dynamic behavior is represented by 
\begin{align}
    \begin{cases}
        \extendedStateVectorDerivative(t) = \stateMatrix{3}\extendedStateVector(t) + \inputVector{3}\totalDisturbanceDerivative(\extendedStateVector,t) + \inputVectorExtendedState{3}\wenchaoInputVectorFieldEstimate\controlSignal(t) \\
        \systemOutput(t) = \outputVector{3}^\top\extendedStateVector(t) + \measurementNoise(t).
    \end{cases}
    \label{eq:extendedStateDynamics}
\end{align}
\begin{remark}
    \label{rem:totalDisturbance}
    Upon Assumptions \ref{ass:1}-\ref{ass:5}, we may claim that total disturbance and its derivative are bounded, i.e.,  $\totalDisturbance\in\ballDomain{\totalDisturbanceBallRadius}$ and $\totalDisturbanceDerivative\in\ballDomain{\totalDisturbanceDerivativeBallRadius}$ for $0<\totalDisturbanceBallRadius,\totalDisturbanceDerivativeBallRadius<\infty$.
\end{remark}

In this work, we will focus on the performance of a closed-loop system designed for stabilization around the equilibrium at point $\stateVectorEquilibrium\triangleq[0 \ 0]^\top$. In \eqref{eq:extendedStateDynamics}, we find that only a first element of the extended state vector can be measured. Thus, to properly design a robust controller described by a generalized formula
\begin{align}
    \controlSignal(\extendedStateVectorEstimate,t) \triangleq \frac{1}{\wenchaoInputVectorFieldEstimate}\left[\feedbackController(\stateVectorEstimate,t) - \totalDisturbanceEstimate(t)\right],
    \label{eq:genericController}
\end{align}
that consists of a stabilizing feedback controller $\feedbackController:\realNumbers^2\times\nonNegativeRealNumbers\rightarrow\realNumbers$ and the feedforward component compensating total disturbance,
we need to estimate the extended state vector $\extendedStateVectorEstimate\triangleq[\stateVectorEstimate^\top \totalDisturbanceEstimate]^\top$. The Luenberger-like extended state observer structure, that is most commonly utilized along the existing literature \cite{madonski2019,xue2017} and has a proven track of industrial applicability \cite{sun2021,sun2005}, has the form
\begin{align}
    \extendedStateVectorEstimateDerivative(t) = \stateMatrix{3}\extendedStateVectorEstimate(t) + \inputVectorExtendedState{3}\wenchaoInputVectorFieldEstimate\controlSignal(\extendedStateVectorEstimate,t) + \observerGainVector\left[\systemOutput(t)-\outputVector{3}^\top\extendedStateVectorEstimate(t)\right],
    \label{eq:esoObserver}
\end{align}
where the observer gain vector $\observerGainVector=[\observerGainVectorElement{1} \ \observerGainVectorElement{2} \ \observerGainVectorElement{3}]^\top\in\realNumbers^3$ is designed such that the state matrix of the estimation error dynamics is Hurwitz. In other words, for the estimation error defined as $\extendedStateVectorObservationError\triangleq\extendedStateVector-\extendedStateVectorEstimate$ which dynamics, derived from \eqref{eq:extendedStateDynamics} and \eqref{eq:esoObserver}, is represented by 
\begin{align}
    \extendedStateVectorObservationErrorDerivative(t) = \underbrace{(\stateMatrix{3}-\observerGainVector\outputVector{3}^\top)}_{\observationErrorStateMatrix}\extendedStateVectorObservationError(t) + \inputVector{3}\totalDisturbanceDerivative(\extendedStateVector,t) - \observerGainVector\measurementNoise(t),
    \label{eq:observationErrorDynamics}
\end{align}
and the matrix $\observationErrorStateMatrix$ is Hurwitz.
The observer gains are calculated using the pole-placement procedure, by selecting the real negative eigenvalues of the state matrix in the closed-loop observation subsystem \eqref{eq:observationErrorDynamics}, i.e.  $\eigenvalue{i}:=\eigenvalue{i}(\observationErrorStateMatrix)\in\negativeRealNumbers$ for $i\in\{1,2,3\}$, implying
\begin{align}
    \observerGainVectorElement{1} &= -(\eigenvalue{1} + \eigenvalue{2} + \eigenvalue{3}) \label{eq:observerTuning1}\\ 
    \observerGainVectorElement{1} &= \eigenvalue{1}\eigenvalue{2} + \eigenvalue{1}\eigenvalue{3} + \eigenvalue{2}\eigenvalue{3} \\ 
    \observerGainVectorElement{3} &= -\eigenvalue{1}\eigenvalue{2}\eigenvalue{3}. \label{eq:observerTuning3}
\end{align}
The chosen gain parametrization allows us to formulate following theorem, which we prove in \ref{app:proofTheorem1}.
\begin{theorem}
    \label{th:1}
    Under Assumptions \ref{ass:1}-\ref{ass:5} and Remark \ref{rem:totalDisturbance}, the observation error subsystem \eqref{eq:observationErrorDynamics}, subject to the observer gain parametrization from \eqref{eq:observerTuning1}-\eqref{eq:observerTuning3},  locally satisfies
    \begin{align}
        \|\extendedStateVectorObservationError(t)\| \leq c_1\|\extendedStateVectorObservationError(0)\|e^{-c_2t} + c_3\left[\totalDisturbanceDerivativeBallRadius + \|\observerGainVector\|\measurementNoiseBallRadius\right],
        \label{eq:th1}
    \end{align}
    for some constants $c_1,c_2,c_3>0$.
\end{theorem}
The result presented in Theorem \ref{th:1} shows that as $t\rightarrow\infty$, the first component of \eqref{eq:th1} resulting from the non-zero initial conditions converges to zero, while the norm of the overall estimation error is bounded around some vicinity of zero dependent on the values $\totalDisturbanceDerivativeBallRadius$ and $\measurementNoiseBallRadius$. Although the general parametrization from \eqref{eq:observerTuning1}-\eqref{eq:observerTuning3} did not allow us to analytically present a direct correlation between the observer gains and the convergence speed or the disturbance rejection impact, result \eqref{eq:th1} proved that an increase of observer gains decrease the estimation robustness with respect to the measurement noise.  

The most commonly utilized bandwidth-parametrization of vector $\observerGainVector$, originally introduced in \cite{gao2006scaling}, locates all eigenvalues of matrix $\observationErrorStateMatrix$ at the same place, i.e., $\eigenvalue{i}=-\observerBandwidth$ for $i\in\{1,2,3\}$, resulting in 
\begin{align}
    \label{eq:bandwidthParametrization}
    \observerGainVector = [3\observerBandwidth \ 3\observerBandwidth^2 \ \observerBandwidth^3]^\top, \quad \observerBandwidth>0.
\end{align}
Such parametrization allows us to provide a stronger analytical result, comparing to the one presented in Theorem \ref{th:1}, formulated within the following theorem, which we prove in \ref{app:proofTheorem2}.

\begin{theorem}
    \label{th:2}
    Under Assumptions \ref{ass:1}-\ref{ass:5} and Remark \ref{rem:totalDisturbance}, the observation error subsystem \eqref{eq:observationErrorDynamics}, subject to the observer gain parametrization from \eqref{eq:bandwidthParametrization}, locally satisfies
    \begin{align}
        \|\extendedStateVectorObservationError(t)\|\leq & \max\{\observerBandwidth^{-2},\observerBandwidth^{2}\}e^{-c_4\observerBandwidth t}c_5\|\extendedStateVectorObservationError(0)\| \nonumber\\
        &+ \max\{\observerBandwidth^{-2},1\}\frac{1}{\observerBandwidth}c_6\left[\totalDisturbanceDerivativeBallRadius + 3\observerBandwidth^3\measurementNoiseBallRadius\right],
        \label{eq:th2}
        \end{align}
    for some constants $c_4,c_5,c_6>0$.
\end{theorem}
The result presented in Theorem~\ref{th:2} is consistent with the result from Theorem~\ref{th:1} in terms of the increase of potential impact that measurement noise has on the estimation quality with increasing $\observerBandwidth$. Additionally, the total disturbance affects the norm of observation error less for higher $\observerBandwidth$, reaching the ideal compensation of total disturbance $\totalDisturbance$ for $\observerBandwidth\rightarrow0$. In the transient stage, the increase of $\observerBandwidth$ provides faster convergence of initial errors towards zero, but may result in the higher values of $\|\extendedStateVectorObservationError(t)\|$ while the observer peaking due to the dependency of the first component of \eqref{eq:th2} on factor $\max\{\observerBandwidth^{-2},\observerBandwidth^{2}\}$. 

A popular choice of the feedback controller, introduced within the general control law \eqref{eq:genericController}, takes the form of a state feedback
\begin{align}
    \feedbackController(\stateVectorEstimate,t) = -\underbrace{[\controllerGainElement{1} \ \controllerGainElement{2}]}_{\pmb{k}}\stateVectorEstimate,
    \label{eq:feedbackController}
\end{align}
where $\controllerGainElement{1},\controllerGainElement{2}$ were chosen in such a way, that matrix  $\stateMatrix{2}-\inputVector{2}\pmb{k}$ is Hurwitz. For the sake of conciseness of presented results, we introduce a parametrization of $\pmb{k}$ in a way to place the obtained eigenvalues $\forall_{i\in\{1,2\}}\eigenvalue{i}(\stateMatrix{2}-\inputVector{2}\pmb{k})=-\controllerParameter$, resulting in 
\begin{align}
    \controllerGainElement{1} = \controllerParameter^2, \label{eq:par0}\\ 
    \controllerGainElement{2} = 2\controllerParameter,
    \label{eq:par1}
\end{align}
for some $k>0$. The closed-loop system, after substitution of \eqref{eq:genericController} and \eqref{eq:feedbackController} into \eqref{eq:rewrittenOriginalSystem}, is given by:
\begin{align}
    \stateVectorDerivative(t) &= \stateMatrix{2}\stateVector(t) + \inputVector{2}\left[-\pmb{k}\stateVectorEstimate(t)-\totalDisturbanceEstimate(t)+\totalDisturbance(t)\right] \nonumber \\ 
    &= (\stateMatrix{2}-\inputVector{2}\pmb{k})\stateVector(t) + \inputVector{2}\begin{bmatrix} \pmb{k} \ 1 \end{bmatrix}\extendedStateVectorObservationError(t).
    \label{eq:closedloop}
\end{align}
The time response of \eqref{eq:closedloop}  satisfies the following theorem which we prove in \ref{app:proofTheorem3}.
\begin{theorem}
    \label{th:3}
    Under Assumptions \ref{ass:1}-\ref{ass:5}, the response of \eqref{eq:rewrittenOriginalSystem} with controller described by \eqref{eq:genericController} and \eqref{eq:feedbackController}, parametrized by \eqref{eq:par0}-\eqref{eq:par1}, locally satisfies
    \begin{align}
        \|\stateVector(t)\|\leq \max\{\controllerParameter^{-1},\controllerParameter\}c_1e^{-c_2\controllerParameter t}\|\stateVector(0)\| + \frac{1}{\controllerParameter}c_3\max\{\controllerParameter^{-1},\controllerParameter\}\sup_{t\geq0}\|\extendedStateVectorObservationError(t)\|
        \label{eq:th3}
    \end{align}
    for some constants $c_1,c_2,c_3>0$.
\end{theorem}
From \eqref{eq:th3}, we can see that the estimation error norm, which value is highly dependent on the appropriate tuning of ESO parameters according to the Theorems \ref{th:1} and \ref{th:2}, is a perturbing component that pulls the state vector $\stateVector$ away from the equilibrium point $\pmb{x}^*$. 

\begin{remark}
    A set-point stabilization of the state $\stateVector$ in a directly specified equilibrium $\pmb{x}^* \triangleq [0 \ 0]^\top$ may seem as a control task with a very limited application range. In fact, the generalized set-point tracking and trajectory tracking control tasks can be treated as a zero-stabilization control problem if system \eqref{eq:originalSystem} express the dynamics of trajectory/set-point tracking error. Examples of such approach were presented in \cite{lakomy2021ejc} and \cite{madonski2019}.
\end{remark}

\section{Benchmark systems}
\label{sec:models} 
Throughout this paper, we will refer to the closed-loop features of the control structure introduced in Section \ref{sec:prelims}, based on the performance of two benchmark systems. 

\subsection{Nonlinear system (NS)}
\label{sec:wenchaoBenchmark}
The first chosen benchmark system is the nonlinear model introduced in \cite{xue2011} that, using the representation of a dynamic system from \eqref{eq:originalSystem}, is defined by 
\begin{align}
\label{eq:wenchao_1}
    \wenchaoDriftVector(\stateVector,t) &= \sin(\wenchaoBenchmarkParameter{1}t)\stateVectorElement{1}+\stateVectorElement{2}^2 \\
\label{eq:wenchao_2}
    \wenchaoInputVectorField(\stateVector) &= \wenchaoBenchmarkParameter{4} + \wenchaoBenchmarkParameter{5}\sin(\wenchaoBenchmarkParameter{6}\stateVectorElement{2}) \\
\label{eq:wenchao_3}
    \externalDisturbance(t) &= \wenchaoBenchmarkParameter{2}\cos(\wenchaoBenchmarkParameter{3}t),
\end{align}
where $a_1,a_2, ..., a_6 \in \mathbb{R}$ represent the system parameters. Such artificial system, due to the multiple nonlinearities of $\wenchaoDriftVector$, $\wenchaoInputVectorField$, and $\externalDisturbance$, is a good representation of the generic model from \eqref{eq:originalSystem}.


\subsection{Manipulator with 1 degree of freedom (M1D)}
\label{sec:M1D}
\begin{figure}[th]
    \centering
    \includegraphics[width=1.0\linewidth]{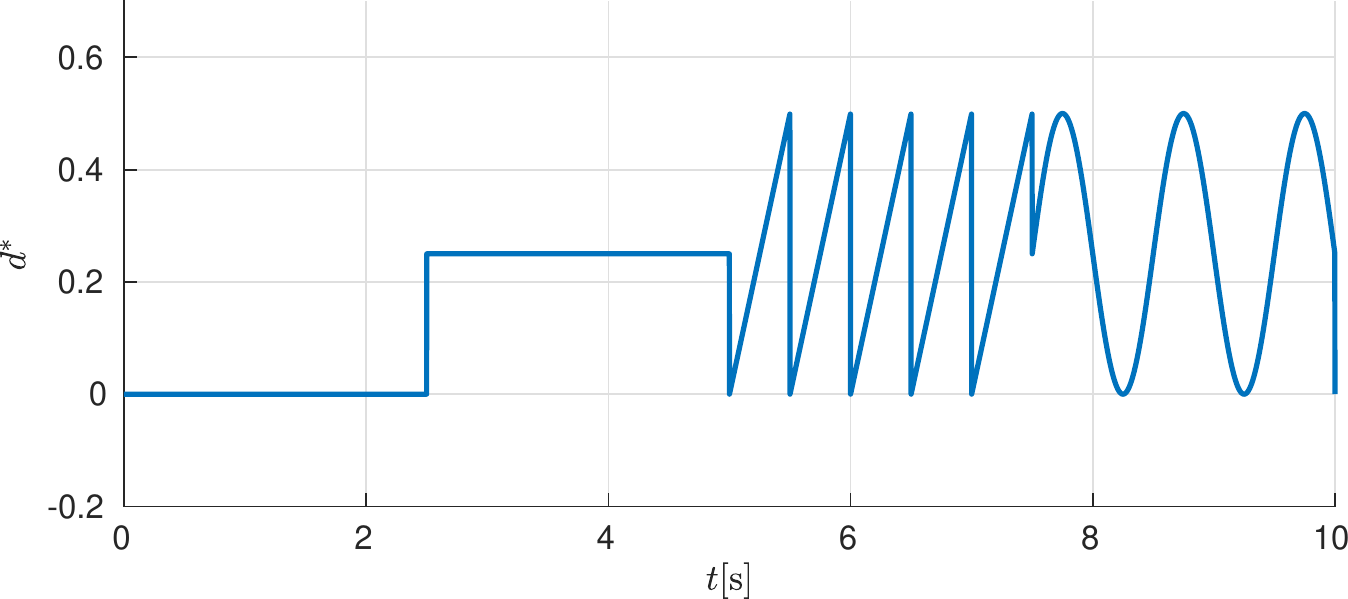}
    \caption{En example of external disturbance \eqref{eq:manipulatorExternalDisturbance} applied to system \eqref{eq:manipulatorDynamics} for parameters $\pmxrBenchmarkParameter{3}=0.25$, $\pmxrBenchmarkParameter{4}=0.25$, $\pmxrBenchmarkParameter{5}=2$, $\pmxrBenchmarkParameter{6}=1$, and $\frac{K_I}{JR}=20.169$ } 
    \label{fig:disturbance}
\end{figure}

A second model utilized within this article considers the motion of a DC-motor driven 1 DoF manipulator, used as a benchmark in \cite{xue2017} and \cite{zhao2021}. The dynamics of the selected object can be represented as
\begin{align}
        \ddot{\theta}(t) = -\frac{RC_b+K_EK_I}{JR}\dot{\theta}(t)-\frac{K_I}{JR}\controlSignal(t)+\externalDisturbance(t),
        \label{eq:manipulatorDynamics}
\end{align}
where $\theta(t)$ is the angular position of the manipulator, $\controlSignal(t)$ corresponds to the reversed voltage put on the DC-motor input, $R$ is the armature resistance, $C_b$ represents the friction coefficient, $K_I$ is a torque constant, $K_E$ is a speed motor constant, and $J$ is the inertia of manipulator. To represent \eqref{eq:manipulatorDynamics} in framework \eqref{eq:originalSystem}, we need to define a state vector $\stateVector\triangleq[\theta \ \dot{\theta}]$, and then, we can write that 
\begin{align}
    \wenchaoDriftVector(\stateVector) &= \pmxrBenchmarkParameter{1}\stateVectorElement{2} \\ 
    \wenchaoInputVectorField &= \pmxrBenchmarkParameter{2},
\end{align}
where $\pmxrBenchmarkParameter{1} = -\frac{RC_b+K_EK_I}{JR}$ and $\pmxrBenchmarkParameter{2} = -\frac{K_I}{JR}$. External disturbance applied to the system is designed as
\begin{align}
    \externalDisturbance(t) &= \begin{cases} 
    0, \ &\textrm{for} \ t\in[0, 2.5) \\ 
    -\frac{K_I}{JR}\pmxrBenchmarkParameter{3}, \ &\textrm{for} \ t\in[2.5,5) \\ 
    -\frac{K_I}{JR}\left[\pmxrBenchmarkParameter{3} + \pmxrBenchmarkParameter{4}\sawtooth(2\pi\pmxrBenchmarkParameter{5}(t-5))\right], \ &\textrm{for} \ t\in[5,7.5) \\ 
    -\frac{K_I}{JR}\left[\pmxrBenchmarkParameter{3} + \pmxrBenchmarkParameter{6}\sin(2\pi\pmxrBenchmarkParameter{7}(t-7.5))\right], \ &\textrm{for} \ t\in[7.5,10) \end{cases},
    \label{eq:manipulatorExternalDisturbance}
\end{align}
where $\pmxrBenchmarkParameter{1}, \pmxrBenchmarkParameter{2},..., \pmxrBenchmarkParameter{7} \in \mathbb{R}$ are system parameters, and $\sawtooth(2\pi\pmxrBenchmarkParameter{5},t-5)$ is a sawtooth function with period equal to $1/\pmxrBenchmarkParameter{5}$.
Function \eqref{eq:manipulatorExternalDisturbance}, visualized in Fig. \ref{fig:disturbance}, consists of a selected set of different function types (constant, sawtooth, and sinusoidal), which could be met during practical operation of manipulator working under varying load. 

\section{Problem formulation}
\label{sec:problem} 

\subsection{Generic ESO tuning task}

The problem considered in this paper is to propose a tuning procedure for the extended state observer \eqref{eq:esoObserver}, with gains parametrized as in \eqref{eq:observerTuning1}-\eqref{eq:observerTuning3}, according to the specified set of user-defined quality criteria. For the bandwidth-parametrization-based tuning introduced in \eqref{eq:bandwidthParametrization} and described in Theorem \ref{th:2}, one can observe that the increase of eigenvalues $\eigenvalue{i}:=\observerBandwidth, \ i\in\{1,2,3\}$ results in (i) faster convergence of the component caused by the initial estimation error $\|\extendedStateVectorObservationError(t)\|$, (ii) reduced maximal value of the estimation error caused by the total disturbance, and (iii) increased impact of the measurement noise on the estimation errors. Although the analysis of the observation error without placing all $\eigenvalue{i}, \ i\in\{1,2,3\}$ at the same point, concluded with Theorem \ref{th:1}, does not present direct analytical relation between the eigenvalues of matrix $\observationErrorStateMatrix$ and the convergence speed and/or the impact of disturbances and measurement noise, it is reasonable to assume that the control performance visible in the simulations may show similar tendencies with decreasing $\eigenvalue{i}$-s and with increasing $\observerBandwidth$ for the bandwidth-parametrized observer. Allowing the eigenvalues of $\observationErrorStateMatrix$ to have different values introduces an additional degree of freedom in the observer design that potentially can improve the desired closed-loop characteristics. 

\subsection{Performance-based ESO tuning task}

To achieve a practically appealing control performance of the observer-based robust control structure, the tuning procedure of ESO parameters should take into account at least several properties, for example, control errors describing the control tracking precision, energy consumption that usually affects the operation costs of the process, and/or the jittering of control signals that can affect the lifetime length of the actuators. 

\begin{figure}[th]
    \centering
    \includegraphics[width=0.9\linewidth]{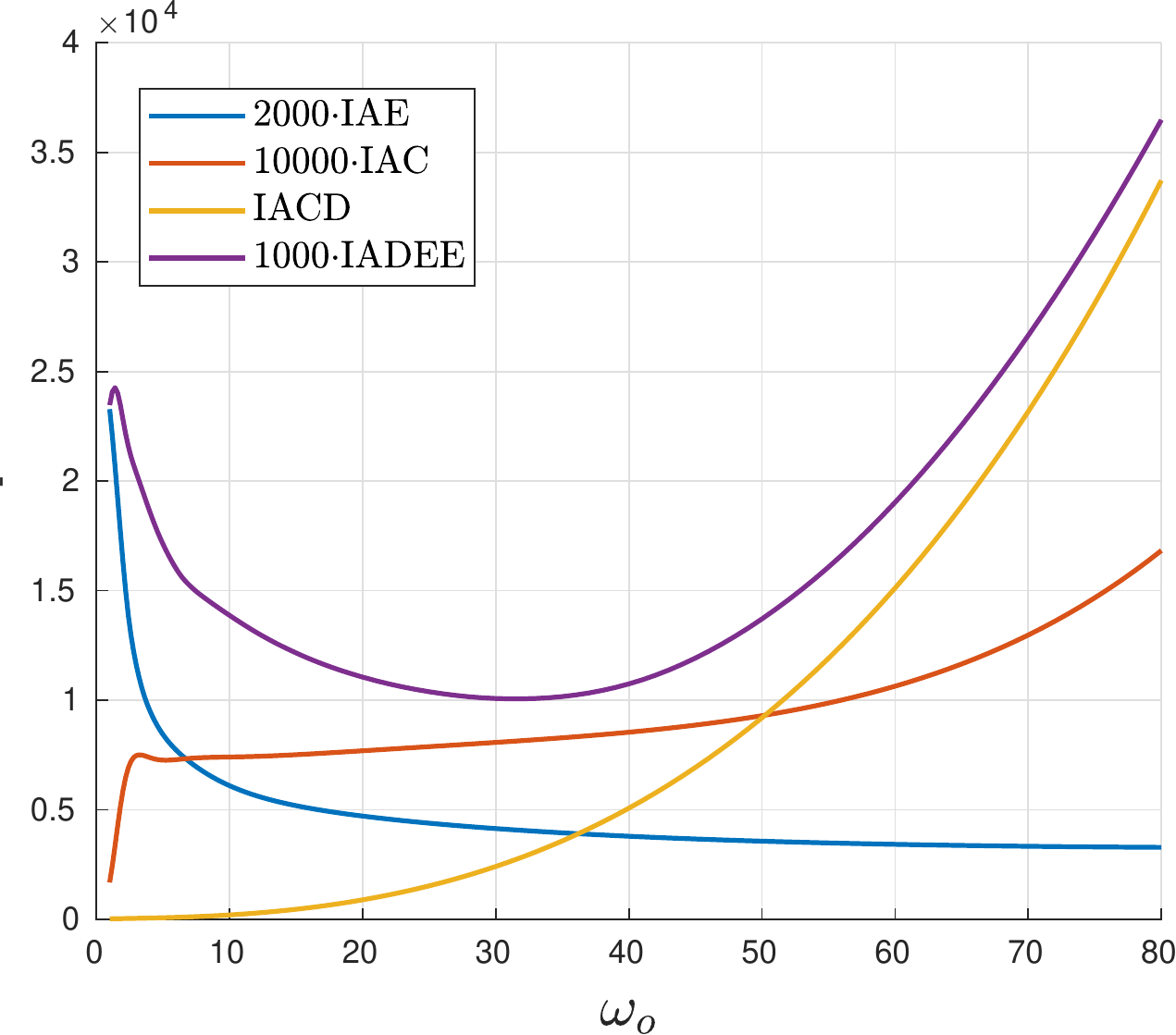}
    \caption{The values of chosen quality criteria for different values of $\observerBandwidth$ parameter} \color{black}
    \label{fig:qualityCriteriaw0}
\end{figure}

Since the robust controller \eqref{eq:genericController} rely directly on the estimation quality of total disturbance, the quality criterion describing the disturbance estimation error  
\begin{align}
\label{eq:IADEE}
    \IADEE = & \int_0^T \left|d(t) - \hat{d}(t)\right| dt
\end{align}
should be the natural selection to consider. Unfortunately, the use of the disturbance observer most frequently means that we cannot calculate the total disturbance, and thus, criterion $\IADEE$ is impossible to calculate. Because of that, we need to rely on other quality criteria. Most frequently, see \cite{zhang2019} and \cite{wang2021}, the observer is tuned in a way to minimize the value of mean control error 
\begin{align}
\label{eq:IAE}
    \IAE = & \int_0^T \left|x_1(t)\right| dt,
\end{align}
which, based on the results presented Theorem \ref{th:3}, should also be affected by the estimation errors. In Fig. \ref{fig:qualityCriteriaw0}, we presented the $\IAE$ criterion obtained for the closed-loop performance of the manipulator described in section \ref{sec:M1D} with the observer parametrized as in \eqref{eq:bandwidthParametrization} and the controller \eqref{eq:feedbackController}  for $\observerBandwidth\in[1,80]$, $k_1 = 16$, $k_2 = 8$, $\wenchaoInputVectorFieldEstimate = -20$, $\pmxrBenchmarkParameter{3}-\wenchaoBenchmarkParameter{6}$ as in the description of Fig. \ref{fig:disturbance}, $\pmxrBenchmarkParameter{1} = -8.8255$, $\pmxrBenchmarkParameter{2} = -20.169$, $\measurementNoiseStandardDeviation=0.0059$, $\stateVector(0) = [1 \ 0]^\top$, and $\extendedStateVectorEstimate(0) = [1 \ 0 \ 0]^\top$. One can see that the $\IAE$ value decreases monotonically with the increase of the observer gains, but the estimation quality expressed with IADEE criterion, and the criteria connected with the mean control signal value and control signal jiterring expressed, respectively, as 
\begin{align}
\label{eq:IAC}
    \IAC = & \int_0^T \left|u(t)\right| dt, \\
\label{eq:IACD}
    \IACD = & \int_0^T \left|\frac{du(t)}{dt}\right| dt
\end{align}
start to grow in the range of large values of $\observerBandwidth$. 

The aim of this work is to propose a tool which
estimates values of $\IADEE$, $\IAE$, $\IAC$ and $\IACD$ criteria for a given control system and a specified set of $\eigenvalue{i}$-s, and allow to select the eigenvalues satisfying $\min_{\eigenvalue{1},\eigenvalue{2},\eigenvalue{3}}\costFunction$, where 
\begin{align}
    \costFunction = \costFunctionWeight{1}\widehat{\IAE}+\costFunctionWeight{2}\widehat{\IAC}+\costFunctionWeight{3}\widehat{\IACD}+\costFunctionWeight{4}\widehat{\IADEE},
    \label{eq:J}
\end{align}
represents the cost function, while $\costFunctionWeight{i} > 0$ are the user-defined values that determine what features should the algorithm prioritize.

\FloatBarrier
\section{Proposed solution}
\label{sec:solution}

\subsection{Proposed ESO tuning method}
\label{sec:tuning_procedure}

There are several ways to approach the problem described in Section \ref{sec:problem}. One can propose tuning rules for some specific types of objects or derive heuristics, which will help an engineer to tune the ESO by hand. However, none of these methods offer a general solution for the ESO tuning for ADRC performance improvement, and they usually require to perform multiple experiments to find satisfactory observer gains.
Disregarding the limitation of the existing tuning methods, let's consider a general form of the solution to the ESO tuning problem -- an \textit{ideal ESO tuner}. Such a tuner should have the following property: given the controlled object, noise parameters, ADRC controller parameters, initial error, and some performance criterion, it calculates the optimal (with the reference to the given criterion) observer gains. In practice, for nontrivial examples, it is hard to define such an \textit{ideal ESO tuner}, because we do not have access to the exact model of the system. However, we conjecture that it is possible to approximate the behavior of such an optimal tuner in a specified range of its parameters. 
f
In order to limit the information needed by the tuner, we propose an alternative formulation that does not need to be aware of the specific performance criterion defined by the control engineer. Thus, instead of searching for an \textit{ideal ESO tuner}, we focus on finding an \textit{ideal performance estimator}. Let's observe, that having a function which for a given controlled object, noise parameters, ADRC controller parameters, initial error, and ESO gains, returns its expected performance in terms of several important criteria, we are able to obtain similar results to the ones obtained with an \textit{ideal ESO tuner}. 
Using an \textit{ideal performance estimator} multiple times with different ESO gains enables to find which one satisfies required properties, for example minimizing some weighted sum of the criteria (see \eqref{eq:J}) or even some hard constraints set on them.
Note, that the range of the reasonable gains to check is bounded based on theoretical considerations presented in Theorem \ref{th:1} and Theorem \ref{th:2}. In this way, we reformulated the problem of finding an ideal ESO tuner, to the problem of first finding a performance estimator and then using a search algorithm to find observer gains that result in satisfactory performance of the control system.

\begin{figure}[th]
    \centering
    \includegraphics[width=\linewidth]{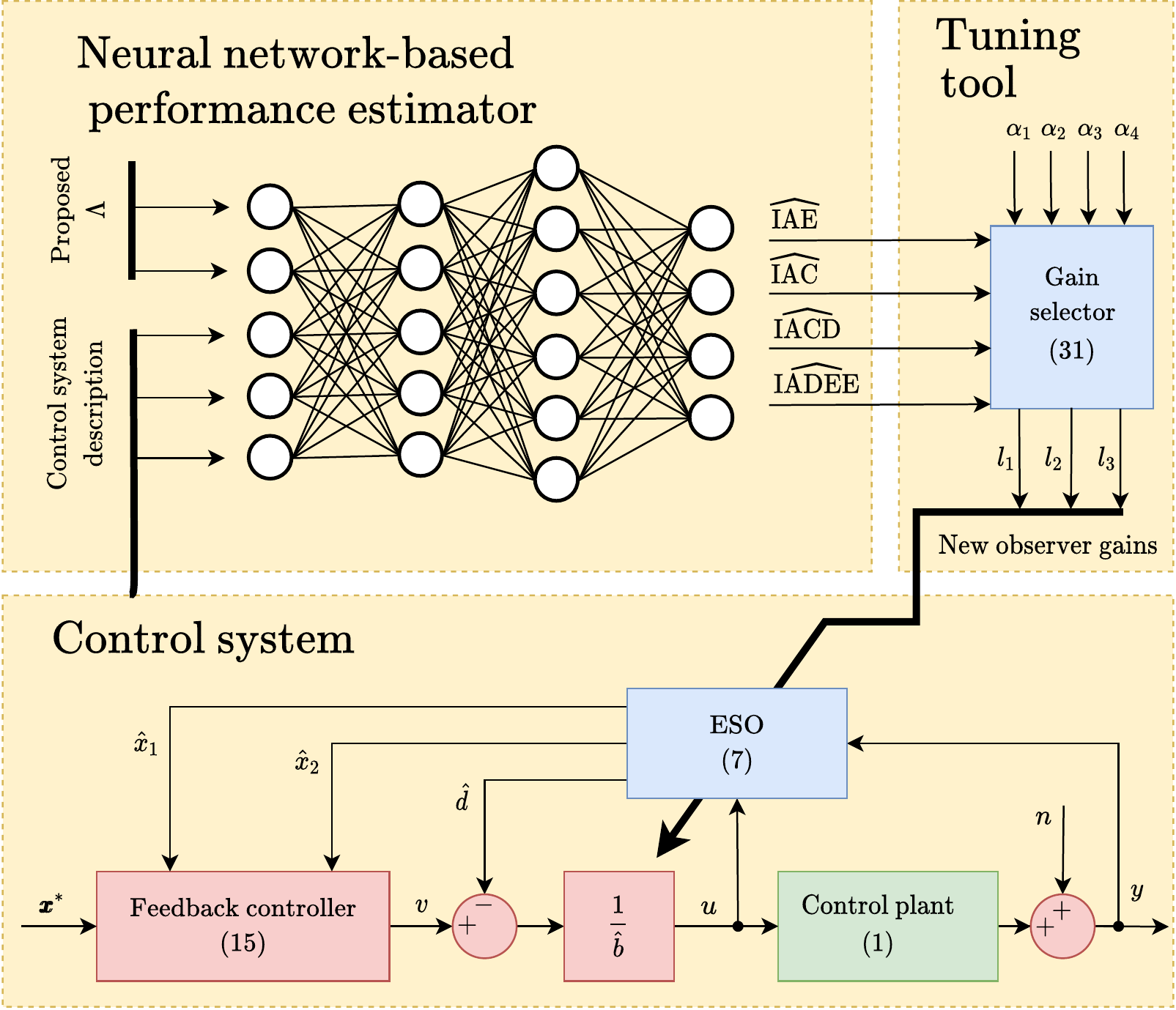}
    \caption{Block diagram of the overall control structure including the NN-based tuning tool.}
    \label{fig:nn_all}
\end{figure}

The general scheme of the proposed neural network-based ESO tuning method is presented in Figure~\ref{fig:nn_all}. Given the description of the control system and multiple candidate sets of the observer gains $\Lambda \triangleq [\lambda_1 \ \lambda_2 \ \lambda_3]^\top$, the neural network estimates the performance criteria values. Based on those estimates, the gain selector selects the gains set, which minimizes user-defined criterion $J$.

\subsection{Neural network-based performance estimator}
In this paper, we will omit the search algorithm part, as there are many appropriate optimization algorithms \cite{random_search, local_greedy_search, genetic_search}, and focus on developing an approximator for the ideal performance estimator. To find such an approximator we follow a data-driven approach and use a neural network, as neural networks are known to be universal approximators with significant representational power, as well as notable generalization properties. The proposed neural network based performance estimator predicts four, in our opinion most representative, integral performance criteria: integral of absolute error (IAE), integral of control effort (IAC), integral of absolute control derivative (IACD), and integral of disturbance estimation error (IADEE), which are defined in \eqref{eq:IADEE}--\eqref{eq:IACD}. However, we presume that the proposed approach will also work for other criteria.

Crucial for the proposed learning system is to feed it with input data, based on which it will be able to determine the values of the aforementioned criteria. 
While it is relatively easy to feed the neural network with the data such as noise standard deviation, ADRC controller parameters, or ESO gains, the core difficulty is to provide a representation of the controlled system.
The simplest idea is to describe it with a set of its parameters, however, such an approach is very inflexible and requires the perfect knowledge of the model structure and identification of its parameters, which can be hard to perform or introduce some additional errors to the tuning procedure. 
Therefore, we describe the system by its transient state estimated by an ESO in some predefined experiment, which we call \textit{basic experiment} or \textit{test experiment}. In this experiment we assume that the system starts from some random initial state $\pmb{x}_{test_0}$, and it is controlled in closed loop by the ADRC based controller defined in \eqref{eq:genericController}, with an ESO defined in \eqref{eq:esoObserver} with the gains set according to bandwidth parametrization (see \eqref{eq:bandwidthParametrization}) with $\omega_0 = 25$. Such a choice of the observer gains seems to be large enough to track the evolution of the system, and small enough to not over-amplify sensor noise, for the objects and noise parameters ranges we consider. As a result, the neural network is fed with a transient of the estimated system extended state $\hat{\pmb{z}} \in \mathbb{R}^{1000 \times 3}$ sampled at $\SI{100}{\hertz}$ through $\SI{10}{\second}$. 

\begin{figure}[th]
    \centering
    \includegraphics[width=0.99\linewidth]{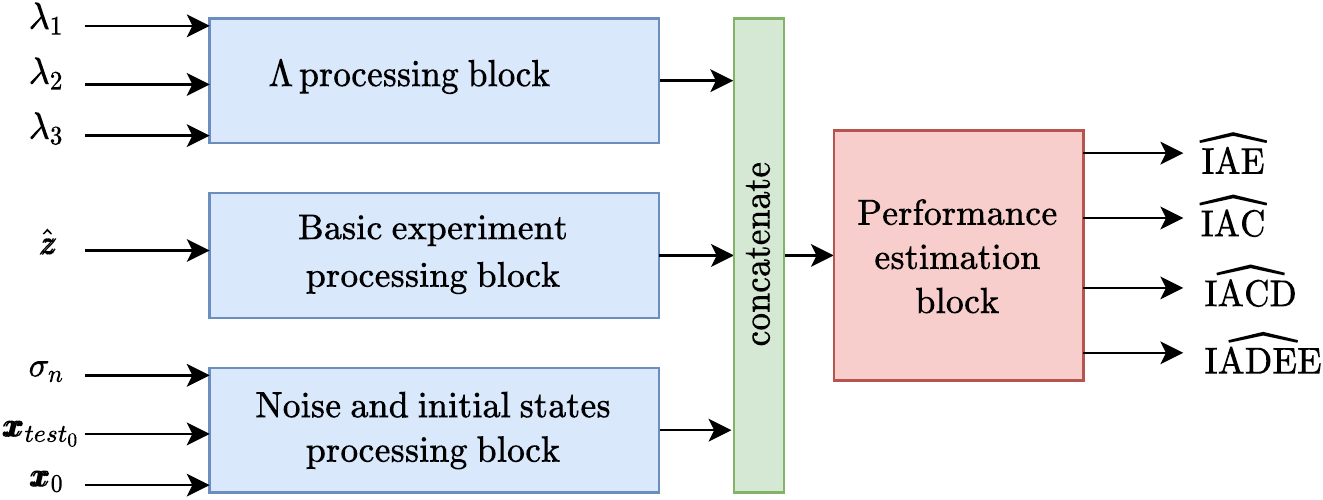}
    \caption{General scheme of the proposed neural network-based performance estimator.}
    \label{fig:nn}
\end{figure}

With both inputs and outputs defined, we now define the neural network architecture. The general scheme of the proposed neural network-based performance estimator is presented in Figure~\ref{fig:nn}. Each of the blocks (except the concatenation operation) is a neural network designed to process different types of data. Here we describe the general purpose of those blocks, as well as their processing structure, inputs, and outputs, while the detailed architectures of those neural networks are presented in the  \ref{sec:appendix_nn}.

The goal of the $\Lambda$ processing block is to create a higher-dimensional representation of the $\lambda_i$ values. An important feature of this processing block is to capture the permutation invariance of the inputs, as the ordering of $\Lambda$ vector has no impact on the observer gains calculated based on them (see \eqref{eq:observerTuning1}--\eqref{eq:observerTuning3}). To achieve that, each of the $\lambda_i$ values is processed separately by a fully connected neural network (FC-NN), and then resultant feature vectors are summed together to achieve permutation invariance built in the architecture.

The Basic experiment processing block creates an embedded representation of the control object from the evolution of the system state estimates during the basic experiment. This signature is crucial in the process of estimation of the performance, as it carries out the information about the specific system controlled by ADRC. The time series of the state estimate is processed by a stack of 1D convolutional neural networks (1D-CNN) in order to extract local features of the signal and then passed to a FC-NN to produce an output feature vector.

The Noise and initial states processing block is used to process other important information about the experiments such as noise standard deviation $\sigma_n$, initial states both in test $\pmb{x}_{test_0}$ and final $\pmb{x}_0$ experiments and create from them another feature vector suitable for further processing. To do so we use standard FC-NN.

Finally, representations built from those three sources of information about the system are concatenated together to form a common feature vector, that contains the entire information needed by the Performance estimator block to predict the values of performance criteria in the final experiment.

\subsection{Datasets}
To use the predictive models, such as the one presented in the previous section, one has to train them using data. In this paper, we train the neural network in a supervised learning paradigm, as the input and output data can be easily gathered from simulations. In this section, we describe the datasets created to train, validate and test our model. All these datasets, for each controlled object type, were created using the same distribution of parameters, but the created dataset elements always appear only in one of them.

For both NS and M1D models, described in Section \ref{sec:models}, the general scheme of gathering data was the same. Firstly, we draw object parameters, noise standard deviation, and initial states for both experiments from the predefined distributions, while setting the rest of the closed-loop system parameters constant. Secondly, we simulate the closed-loop system using a predefined set of observer gains $\lambda_1 = \lambda_2 = \lambda_3 = -25$, from the state $\pmb{x}_{test_0}$, and save the trajectory of the estimated extended state of the system $\hat{\pmb{z}}$. Then, we draw new $\lambda_i$ values from the predefined range and simulate the system once again, from the state $\pmb{x}_0$, and save the resultant performance criteria values. Finally, we save all artifacts from those experiments which are necessary for neural network training.

While the general procedure of data generation is the same for both NS and M1D objects, the ranges of their parameters resulting from the given structure of the control plant are different. In Table \ref{tab:params} we present the ranges of parameters from which we uniformly draw samples to create the NS and M1D datasets. For the definitions of those parameters, please see the detailed descriptions of those objects in Section \ref{sec:prelims}.
Besides those, there are some common constants such as sampling and noise frequency both equal to $\SI{1}{\kilo\hertz}$ and simulation time equal to $\SI{10}{\second}$.

\begin{table}[!htb]
    \caption{Parameters used to create (a) NS and (b) M1D datasets.}
    \label{tab:params}
    \begin{subtable}{.5\linewidth}
      \centering
        \caption{}
        \begin{tabular}{|c|c|}
        \hline
        Parameter & Value\\
        \hline
        $\lambda_i$ & [-80; -1] \\
        $a_1$ & [0; 2]\\
        $a_2$ & [0; 1]\\
        $a_3$ & [0; 2]\\
        $a_4$ & [0.5; 1.5]\\
        $a_5$ & [0; 0.3]\\
        $a_6$ & [0; 2]\\
        $\pmb{x}_{test_0}$ & $[-1; 1]^2$\\
        $\pmb{x}_0$ & $[-1; 1]^2$\\
        $\sigma_n$ & [0; 0.01]\\
        $\wenchaoInputVectorFieldEstimate$ & -1\\
        \hline
        \end{tabular}
    \end{subtable}%
    \begin{subtable}{.5\linewidth}
      \centering
        \caption{}
        \begin{tabular}{|c|c|}
        \hline
        Parameter & Value\\
        \hline
        $\lambda_i$ & [-80; -1] \\
        $\pmxrBenchmarkParameter{1}$ & [-20; -4]\\
        $\pmxrBenchmarkParameter{2}$ &  [-30; -10]\\
        $\pmxrBenchmarkParameter{3}$ & [0; 0.5]\\
        $\pmxrBenchmarkParameter{4}$ & [0; 0.5]\\
        $\pmxrBenchmarkParameter{5}$ & [0; 2]\\
        $\pmxrBenchmarkParameter{6}$ & [0; 0.5]\\
        $\pmxrBenchmarkParameter{7}$ & [0; 2]\\

        $\sigma_n$ & [0; 0.02]\\
        $\pmb{x}_{test_0}$ & $[-\pi; \pi]^2$\\
        $\pmb{x}_0$ & $[-\pi; \pi]^2$\\
        $\wenchaoInputVectorFieldEstimate$ & -20\\
        \hline
        \end{tabular}
    \end{subtable} 
\end{table}

Using this procedure, we created three datasets for each object type: (i) training set with 80000 samples, (ii) validation set with 20000 samples, and (iii) test set with 12000 samples. The training set was used to train the neural network, validation set was used to control the training process and limit overfitting to training data, whereas the test set contains \textit{fresh data}, not used in the training process, to evaluate the performance of the predictive model. 

\begin{figure}[th]
    \centering
    \includegraphics[width=0.99\linewidth]{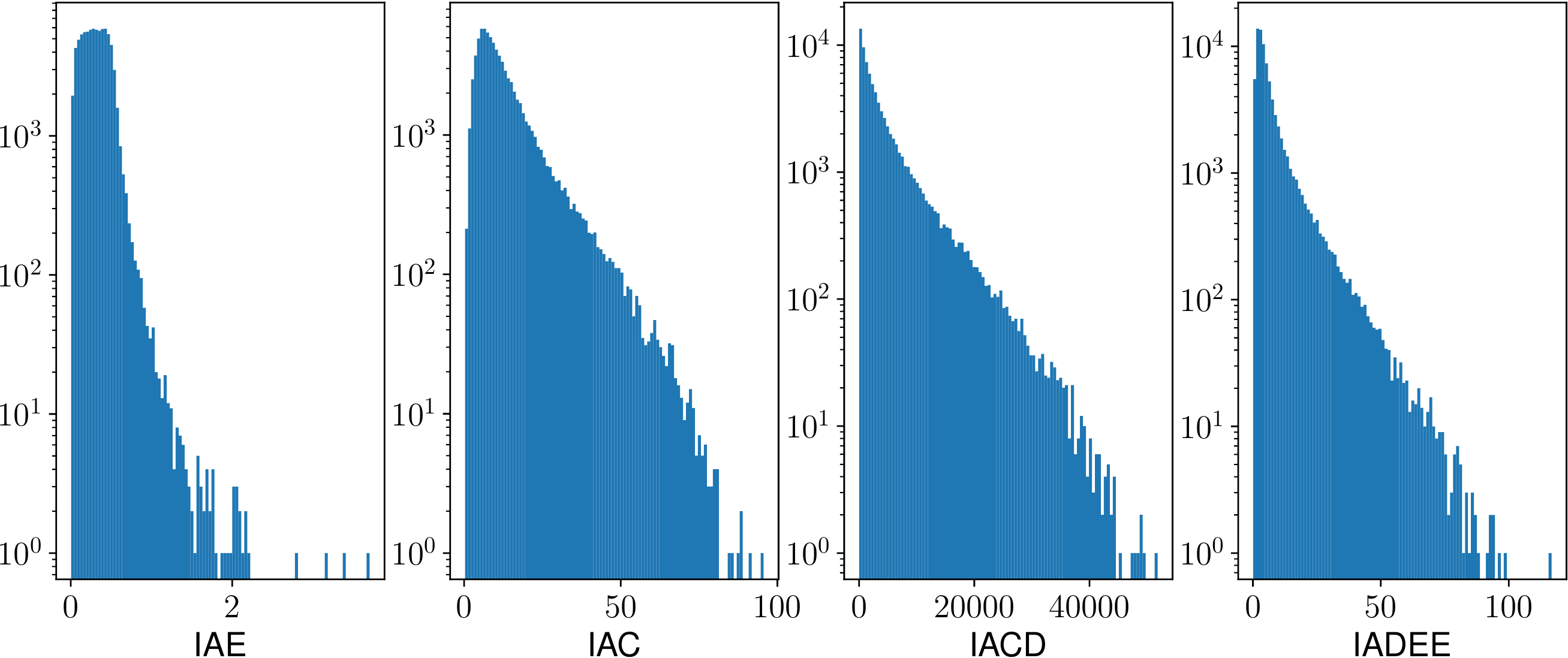}
    \caption{Histograms of the criteria values from the training part of the NS dataset.}
    \label{fig:wenchao_ds}
\end{figure}

\begin{figure}[th]
    \centering
    \includegraphics[width=0.99\linewidth]{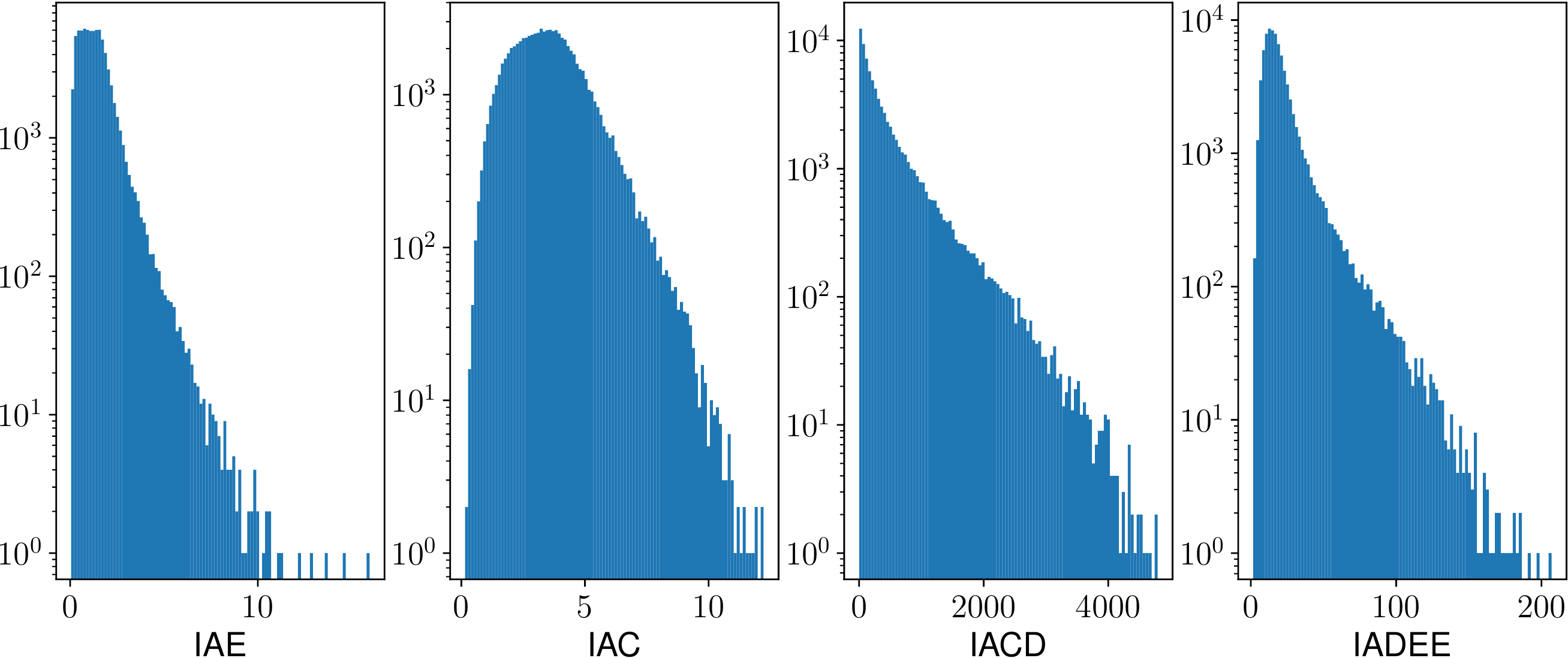}
    \caption{Histograms of the criteria values from the training part of the M1D dataset.}
    \label{fig:M1D_ds}
\end{figure}

In Figures~\ref{fig:wenchao_ds} and \ref{fig:M1D_ds}, we present the histograms of the performance criteria both in NS and M1D training sets. It can be seen that for both datasets the shapes of the criteria distributions are similar. Most of the experiments result in lower values of the criteria, and the number of the outliers is relatively small. Based on these observations, we apply a more aggressive normalization of the neural network outputs, as in the end, for the tuning purposes accurate estimation of the highest values of the performance criteria is not so important.

\section{Results}
\label{sec:experiments}

In order to experimentally validate the approach presented in Section \ref{sec:solution}, we trained the proposed performance estimation neural network using Adam optimizer \cite{adam} with a learning rate of $10^{-4}$, and a batch size of 128 on both M1D and NS datasets. The models were trained for 100 epochs, and the models with the best performance on the validation set were chosen for further tests. It took about an hour to train each of the models using NVIDIA GTX1660 GPU.

It is important to note that before feeding the data to the network we normalized them into $[0; 1]$ range, according to the ranges listed in Table \ref{tab:params} by a linear transformation. Similarly, we normalized the output performance criteria based on the dataset statistics presented in Figures~\ref{fig:wenchao_ds} and \ref{fig:M1D_ds}. As the distributions of the criteria differ between M1D and NS datasets, we used different normalization formulas. For NS dataset: (i) IAE was divided by 3.7, (ii) IAC was divided by 80, (iii) IACD was transformed using following expression $\textrm{IACD}_{norm} = (\log(\textrm{IACD} - 4) - 3) / 8$, and (iv) IADEE was divided by 50.
Similarly, for the M1D dataset: (i) IAE was divided by 8, (ii) IAC was divided by 10, (iii) IACD was divided by 4000, and (iv) IADEE was divided by 100.
For both datasets, the resulting values were finally clipped to the $[0; 1]$ range. While such an operation may lead to some distortions in the case of extremely high values of the criteria, they can be neglected as typically in the tuning procedure highest values of any of the aforementioned criteria are undesirable, as they lead to pathological behaviors, such as extremely high errors or control effort. 

\subsection{Performance estimation}
Firstly, we validate the performance estimation capabilities of the proposed neural networks, as a precise performance estimation leads to the accurate tuning of the ESO gains.
In Table \ref{tab:accuracy} we present the mean absolute errors (MAE) as well as mean absolute percentage errors (MAPE) obtained by our models on the test set (this part of data was not used for training nor model selection), for all considered criteria. Presented results show that the proposed neural networks can estimate each criterion in both datasets with reasonable accuracy, however, the worst results are obtained for IADEE.

\begin{figure}[th]
    \centering
    \includegraphics[width=0.99\linewidth]{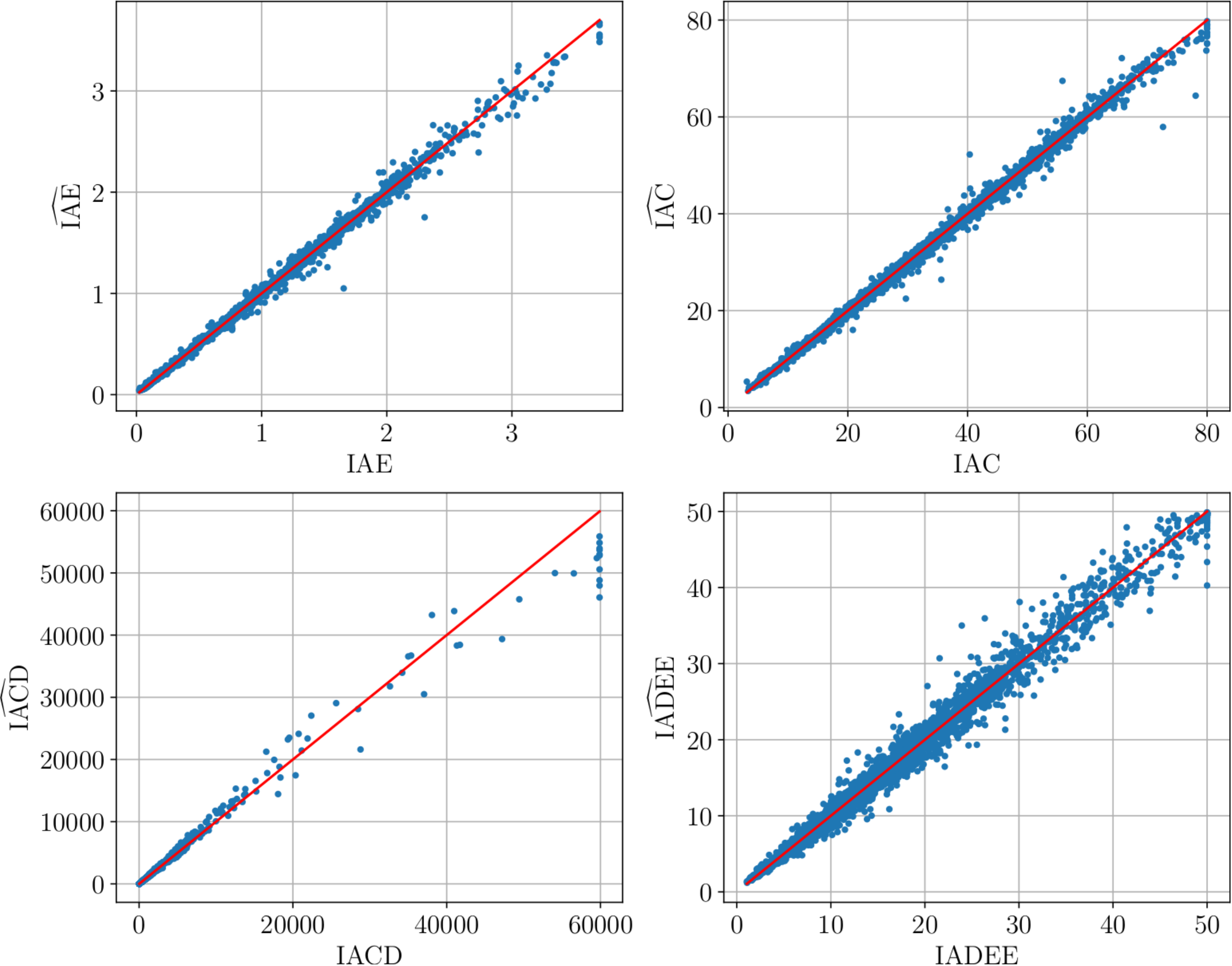}
    \caption{Citeria values predicted by the neural network ploted against the true values form the M1D test set. Red lines denotes the ideal estimator performance ($y = x$).}
    \label{fig:M1D_errors}
\end{figure}

\begin{figure}[th]
    \centering
    \includegraphics[width=0.99\linewidth]{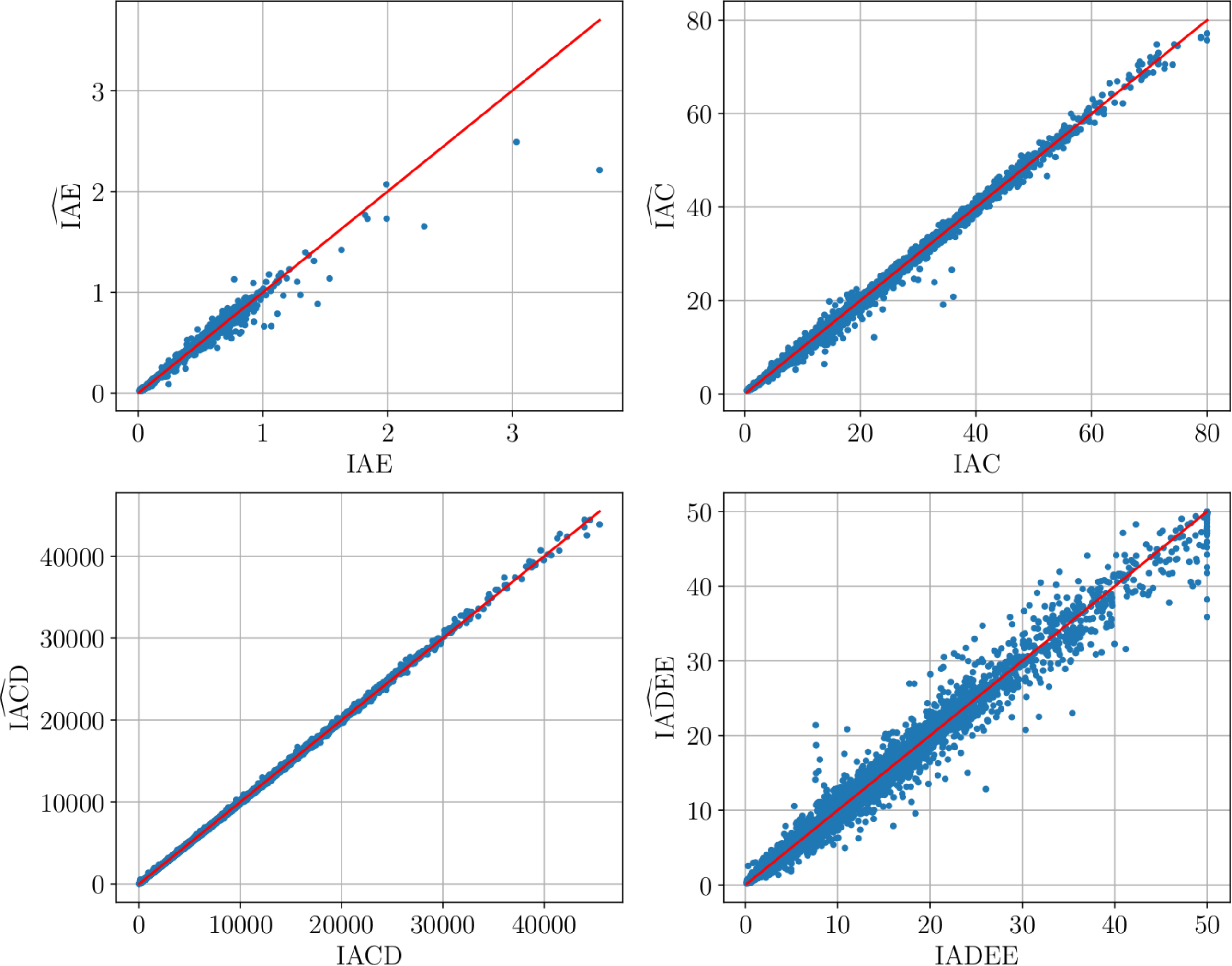}
    \caption{Citeria values predicted by the neural network ploted against the true values form the NS test set. Red lines denotes the ideal estimator performance ($y = x$).}
    \label{fig:wenchao_errors}
\end{figure}

For a more detailed view of the errors made by the performance estimators, one can analyze the plots in Figures \ref{fig:M1D_errors} and \ref{fig:wenchao_errors}. In those charts we present predicted values for all criteria plotted against their true values. The red line visualizes the ideal estimator performance. One can see, that the obtained results, in general, lie very close to the red lines. The thinnest plots almost without outliers are obtained for IACD, whereas thickest for IADEE. In the case of IAE and IAC, there are some outliers, however, the main mass of the predictions lies in the close neighborhood of the ideal estimator line.



\begin{table*}[!th]
\centering
\caption{Mean absolute errors (MAE) and mean absolute percentage errors (MAPE) on the test sets, for all considered criteria. Mean for MAE was omitted due to the different units of each criterion.}
\vspace{0.3cm}
\begin{tabular}{c|cc|cc}
 & \multicolumn{2}{c}{M1D} & \multicolumn{2}{c}{NS}\\
Criterion & MAE & MAPE & MAE & MAPE\\
\hline
IAE & 0.02 $\pm$ 0.04 & 1.74 $\pm$ 3.48 & 0.01 $\pm$ 0.02 & 2.48 $\pm$ 6.47\\
IAC & 0.05 $\pm$ 0.06 & 1.38 $\pm$ 1.68 & 0.36 $\pm$ 0.5 & 3.15 $\pm$ 3.82\\
IACD & 6.11 $\pm$ 9.42 & 3.15 $\pm$ 12.46 & 52.21 $\pm$ 87.95 & 1.88 $\pm$ 5.90 \\
IADEE & 0.79 $\pm$ 1.27 & 3.43 $\pm$ 3.72 & 0.53 $\pm$ 0.89 & 8.40 $\pm$ 12.29\\
\hline
MEAN & & \textbf{ 2.43 $\pm$ 6.84} & &  \textbf{3.98 $\pm$ 8.20}\\
\end{tabular}
\label{tab:accuracy}
\end{table*}

\subsection{Neural network based ESO tuning}
Results obtained in the previous experiments show that the proposed approach can be used to estimate the performance of the closed-loop system with reasonable accuracy. In the following experiments, we validate whether the ESO tuning method utilizing the proposed performance estimators is able to propose reasonable observer gains, and lead to the near-optimal performance in terms of the given criterion. 
Moreover, we present how mixing the IAE, IAC, IACD, and IADEE criteria leads to different behaviors of the tuned system.
As guessing the $\omega_0$ is a popular way of tuning ESO gains, we use random values of $\omega_0 \in [-1; -80]$ as a baseline for our tuning method.

Firstly, we show the results of the proposed tuning procedure for a few example systems. The tuning procedure with the use of a neural network-based performance estimator is described in detail in Section \ref{sec:tuning_procedure}. Here, we are searching for the best ESO gains using a three dimensional uniform grid of the eigenvalues of the state matrix in the closed-loop observation subsystem~\eqref{eq:observationErrorDynamics}: $\Lambda \in \{[-1 - 79s, \ -1 - 79t, \ -1 - 79u]^\top \quad | \quad s, t, u\in\{0, \frac{1}{20}, ..., 1\}\}$.

We first, consider an NS system \eqref{eq:wenchao_1}--\eqref{eq:wenchao_3} with the model parameters $(\wenchaoBenchmarkParameter{1}, \wenchaoBenchmarkParameter{2}, \ldots, \wenchaoBenchmarkParameter{7}) = (1, 0.5, 1, 1, 0.15, 1)$, noise standard deviation $\sigma_n = 0.007$, initial state in the basic experiment run $\pmb{x}_{test_{0}} = [0.2 \quad 0.1]^T$ and initial state in the target run $\pmb{x}_0 = [-0.9 \quad -0.5]^T$, which was not present in any dataset used for training.
In Table \ref{tab:wenchao_tuning} we present several sets of $\Lambda$ values chosen with the use of our tuning solution for a few different performance criteria, together with the particular criterion values. For reference, we also reported the smallest possible criterion value with corresponding $\Lambda$ in general, and for the bandwidth parametrization (for the chosen quantization of $\lambda_i$ values). One can see that for all tested criteria our proposed solution achieves near-optimal performance. Similarly, bandwidth parametrization allows for finding a near-optimal solution for all tested criteria definitions, however, the optimal solutions are found for two $\lambda_i$ values equal, while the third may be different.

\begin{table*}[!th]
\footnotesize
\renewcommand{\arraystretch}{1.6}
\centering
\caption{Comparison of the tuning results achieved by the proposed solution (NN-based estimator) with the best achievable tuning in general (Ideal estimator) and using bandwidth parametrization, predicts near-optimal $\Lambda$ for all considered criteria.}
\vspace{0.3cm}
\begin{tabular}{cccc|cc|cc|cc} \hline \hline
\multicolumn{4}{c}{Criteria weights} & \multicolumn{2}{c}{\makecell{NN-based\\ estimator}} & \multicolumn{2}{c}{\makecell{Ideal\\ estimator}} & \multicolumn{2}{c}{\makecell{Bandwidth\\ parametrization}}\\ \hline
$\alpha_1$ & $\alpha_2$ & $\alpha_3$ & $\alpha_4$ & $J$ & $\Lambda$ & $J$ & $\Lambda$ & $J$ & $\observerBandwidth$ \\
\hline
10 & 1 & 0 & 0 & 7.01 & [-17.6, -17.6, -13.5] & 6.99 & [-25.9, -13.5, -13.5] & 7.04 & -17.6\\
\hline
20 & 1 & 0.005 & 0 & 11.46 & [-13.5, -13.5,  -9.3] & 11.46 & [-13.5 -13.5  -9.3] & 11.48 & -13.5 \\
\hline
50 & 0 & 0 & 0.1 & 7.7 & [-38.4, -38.4, -34.3] & 7.54 & [-34.3, -34.3, -34.3] & 7.54 & -34.3 \\ \hline \hline
\end{tabular}
\label{tab:wenchao_tuning}
\end{table*}

\begin{figure*}[h!]
    \centering
    \includegraphics[width=0.8\linewidth]{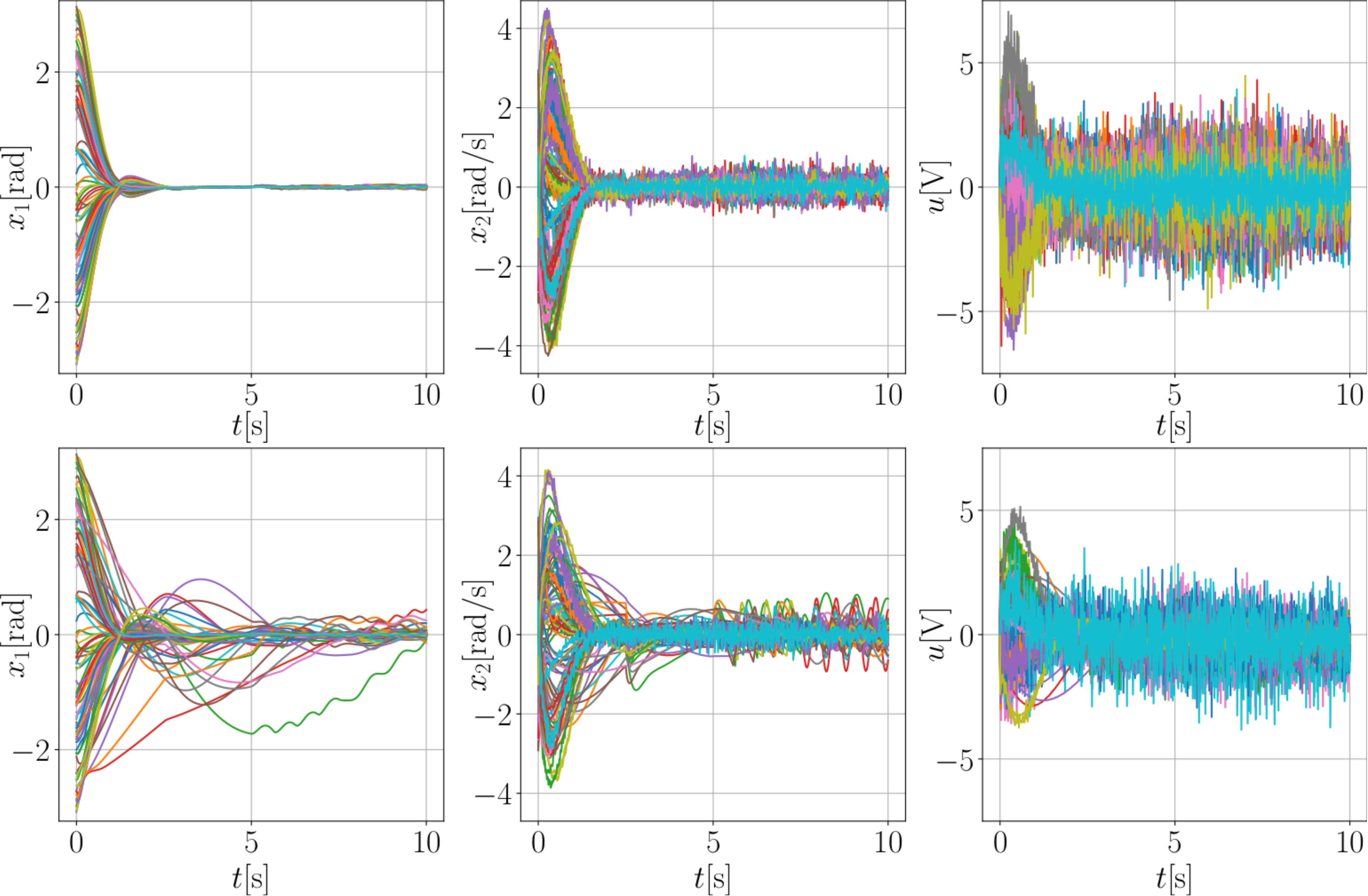}
    \caption{Transient states for a hundred of random M1D objects. Top row contains transients for ADRC with ESO tuned with the proposed method for the performance criterion $J = \widehat{IAE}$, while bottom with random $\omega_0$ value.}
    \label{fig:M1D_1_IAE}
\end{figure*}

\begin{figure*}[h!]
    \centering
    \includegraphics[width=0.84\linewidth]{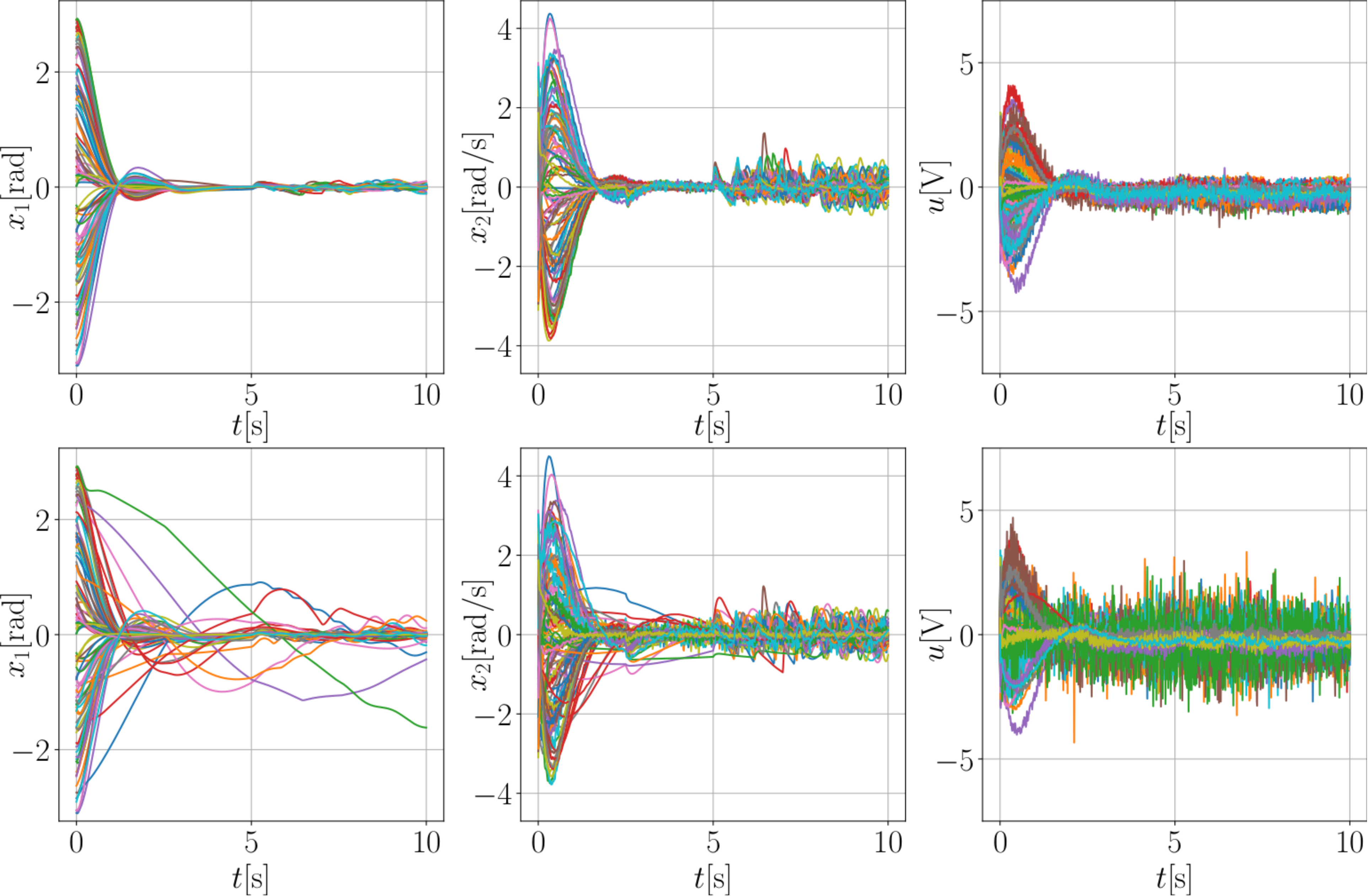}
    \caption{Transient states for a hundred of random M1D objects. Top row contains transients for ADRC with ESO tuned with the proposed method for the performance criterion $J = \widehat{\IAE} + \widehat{\IAC}$, while bottom with random $\omega_0$ value.}
    \label{fig:M1D_0_5_IAE_0_5_IAC}
\end{figure*}

\begin{figure*}[h!]
    \centering
    \includegraphics[width=0.84\linewidth]{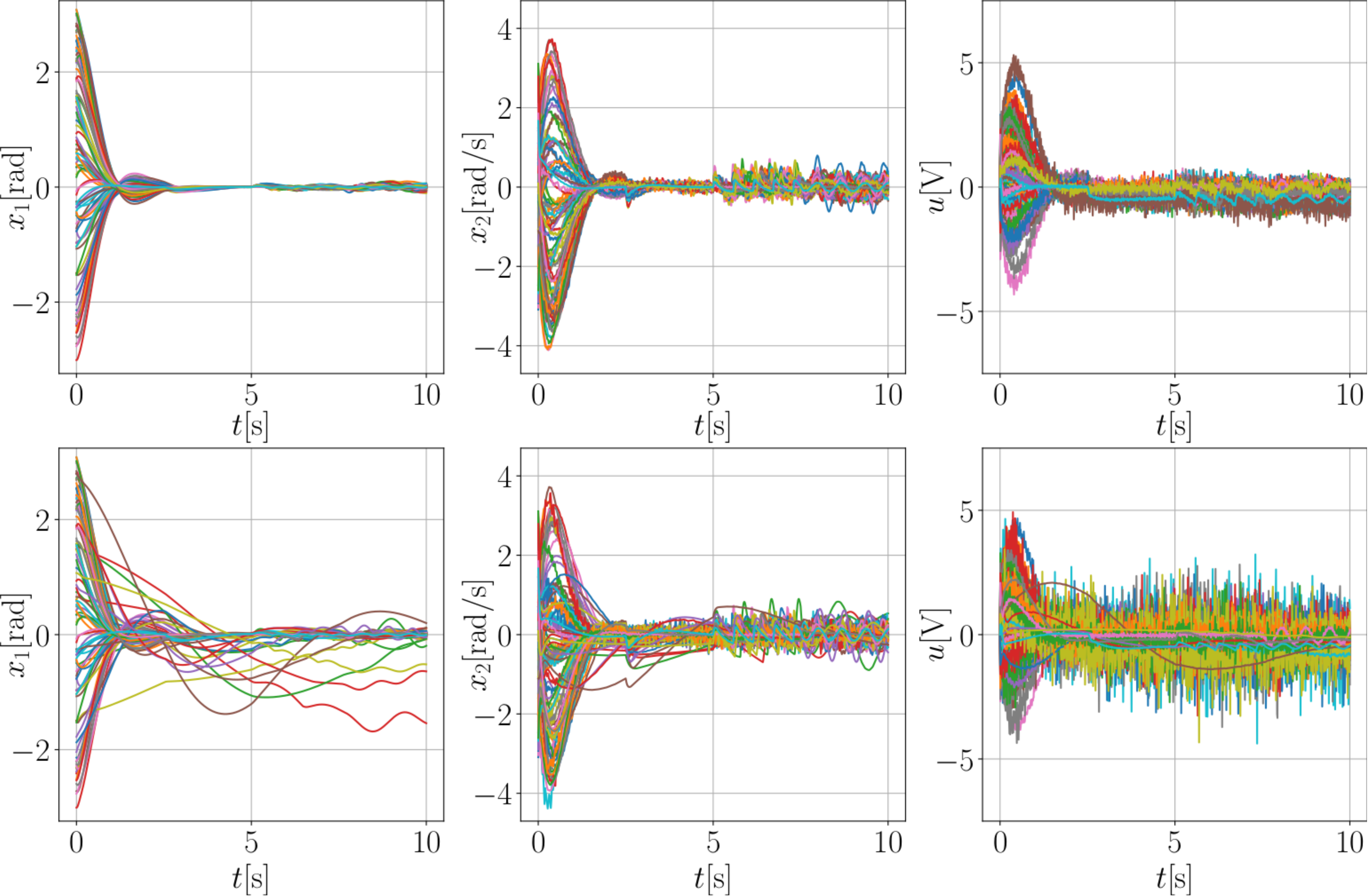}
    \caption{Transient states for a hundred of random M1D objects. Top row contains transients for ADRC with ESO tuned with the proposed method for the performance criterion $J = 0.9 \cdot \widehat{\IAE} + 0.0998 \cdot \widehat{\IAC} + 0.0002 \cdot \widehat{\IACD}$, while bottom with random $\omega_0$ value.}
    \label{fig:M1D_0_9_IAE_0_0998_IAC_0_0002_IACD}
\end{figure*}

\begin{figure*}[h!]
    \centering
    \includegraphics[width=0.8\linewidth]{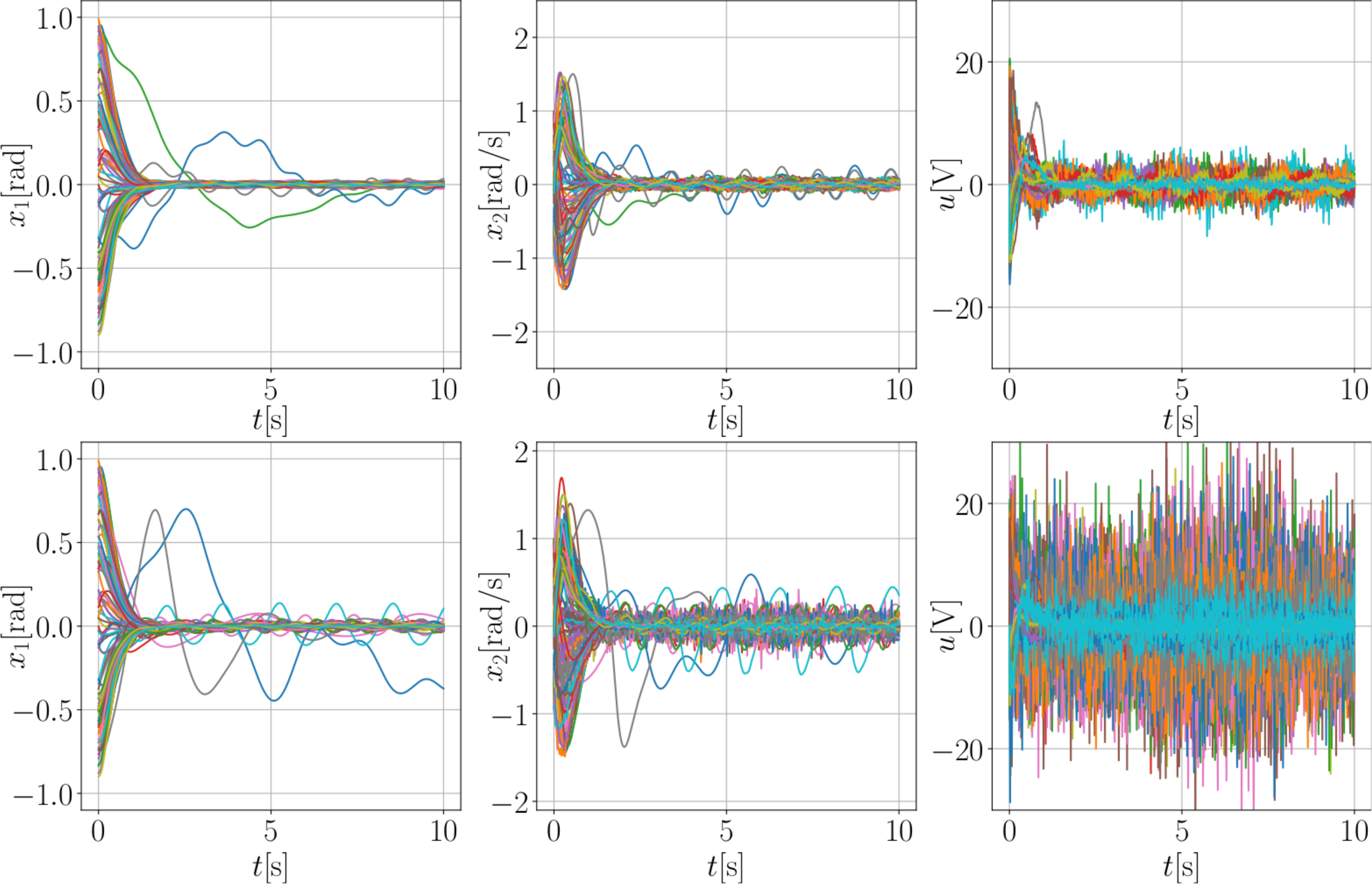}
    \caption{Transient states for a hundred of random NS objects. Top row contains transients for ADRC with ESO tuned with the proposed method for the performance criterion IADEE, while bottom with random $\omega_0$ value.}
    \label{fig:wenchao_IADEE_1}
\end{figure*}

Next, we evaluate the influence of the tuning on the control quality by simulating a hundred random models with the ESO gains selected with the proposed method and compare their transients with the ones obtained for observers with randomly chosen $\omega_0$. Results of those comparisons, for four different performance criteria, are presented in Figures~\ref{fig:M1D_1_IAE}, \ref{fig:M1D_0_5_IAE_0_5_IAC}, \ref{fig:M1D_0_9_IAE_0_0998_IAC_0_0002_IACD} and \ref{fig:wenchao_IADEE_1}. It is clearly visible in Figure~\ref{fig:M1D_1_IAE}, that optimizing only for IAE ($\alpha_1 = 1, \alpha_2=\alpha_3=\alpha_4=0$), which is a popular control objective \cite{zhang2019, wang2021}, with the use of the proposed method is possible and effective. Even though $x_1$ is driven to the close vicinity of zero very quickly, this requires an enormous control effort, which is presented in the rightmost plot. In order to limit that undesirable behavior, we choose a different criterion IAE + IAC ($\alpha_1 = \alpha_2 = 1, \alpha_3=\alpha_4=0$), and perform the same experiment, which results are presented in Figure~\ref{fig:M1D_0_5_IAE_0_5_IAC}. We can observe a slight decrease in the $x_1$ tracking but at the same time a strong improvement in the case of the absolute values of the control signal $u$. In order to also limit the sudden changes of the control signal, we add IACD to the criterion ($\alpha_1 = 0.9, \alpha_2 = 0.0998, \alpha_3=0.0002, \alpha_4=0$) and present the results of this experiment in Figure~\ref{fig:M1D_0_9_IAE_0_0998_IAC_0_0002_IACD}. Without lowering the $x_1$ and $u$ signals suppression performance, we were able to obtain a much smoother control signal, which is beneficial for the controller life span.

In Figure~\ref{fig:wenchao_IADEE_1} we present a tuning performance of our method with a very different criterion -- IADEE only, for a different set of control objects -- NS. It is interesting that disregarding crucial performance metrics like IAE or IAC, and optimizing gains only for the tracking of the total disturbance, we obtain a reasonable quality of $x_1$ suppression (in almost all the cases), as well as limited and smoothed control signal $u$. Surprisingly, if one has no idea what kind of criterion suits the best for its purposes, the choice of IADEE appears to be a safe initial guess.

Obtained results show that by using our tuning method one can greatly improve the behavior of the ADRC control with a very low effort, as the only thing one needs to do is to perform a single closed-loop experiment with the reasonable ESO gains calculated for the $\lambda_1 = \lambda_2 = \lambda_3 = -25$.
Moreover, presented results show how the construction of the criterion affects the behaviors of the controlled system.
In those experiments, we considered previously unseen models drawn from the distribution on which the neural network based performance estimators were trained, however, it may not be the case in real-world applications. 
Therefore, we will next focus on the much harder cases where neural networks will be working outside of the training data distributions.

\subsection{Generalization abilities}
\begin{figure*}[h!]
    \centering
    \includegraphics[width=0.85\linewidth]{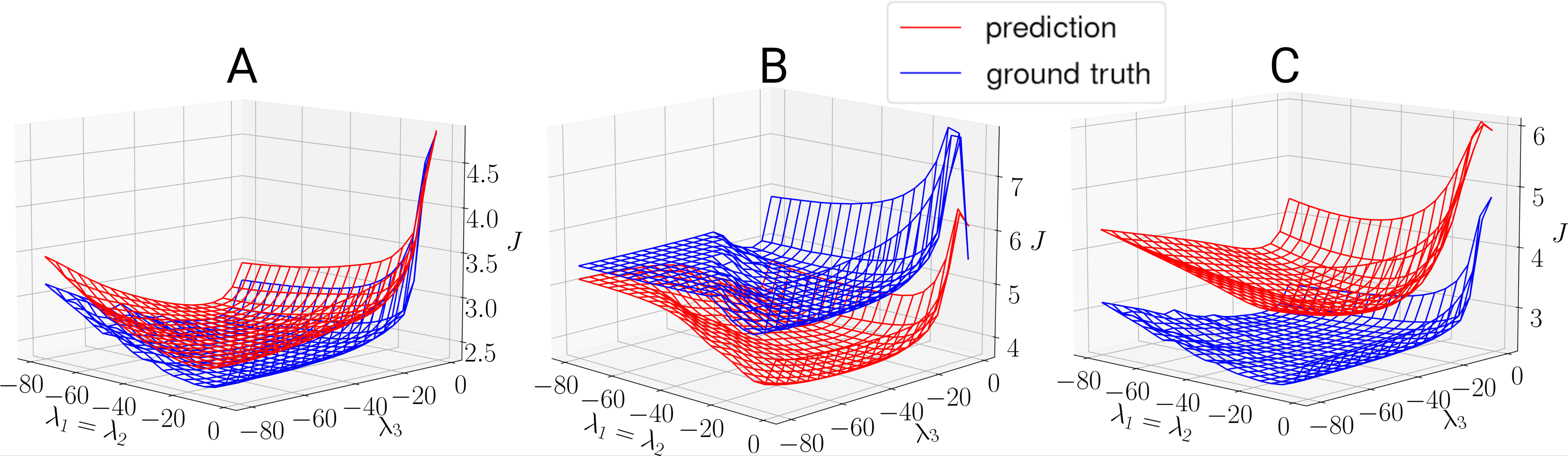}
    \caption{Performance landscapes obtained for 3 setups: A) reference M1D model from the distribution of the training set (however not included in it); B) M1D model from outside of the training set distribution; C) reference M1D model, however, the predictor was trained only on the NS dataset. In all 3 cases, the shape of the predicted surfaces resembles the ground truth, which makes it possible to use the proposed predictor for ADRC tuning even for the object from outside of the training set. }
    \label{fig:generalization}
\end{figure*}

In order to show the generalization abilities of the proposed ESO tuning solution we tested the neural network based performance estimators in some challenging setups, which do not belong to the distribution of the training, validation and test sets. As a reference we use a M1D model with the following parameters: $(\pmxrBenchmarkParameter{1}, \pmxrBenchmarkParameter{2}, \ldots, \pmxrBenchmarkParameter{7}) = (-12, -20, 0.5, 0.5, 1, 0.5, 1)$, noise standard deviation $\sigma_n = 0.01$, initial state in the basic experiment run $\pmb{x}_{test_{0}} = [-0.5 \quad -0.5]^\top$ and initial state in the target run $\pmb{x}_0 = [0.5 \quad 0.5]^\top$. As an example of the M1D model from outside of the training set distribution we choose M1D with: $(\pmxrBenchmarkParameter{1}, \pmxrBenchmarkParameter{2}, \ldots, \pmxrBenchmarkParameter{7}) = (-30, -40, 1, 1, 3, 1, 3)$, noise standard deviation $\sigma_n = 0.03$, initial state in the basic experiment run $\pmb{x}_{test_{0}} = [0.5 \quad 0.5]^\top$ and initial state in the target run $\pmb{x}_0 = [1 \quad 1]^\top$.
To present the generalization power of the proposed approach we plot a ground truth model performance criterion, as well as, predicted criterion value, depending on the chosen values of $\lambda_1, \lambda_2, \lambda_3$. 
For visualization purposes, we assume that $\lambda_3 = \lambda_2$, which allows us to plot the resultant performance landscapes in 3D. For this experiment, we chose the criterion to be an arithmetic mean of the IAE and IAC. Performance surfaces are presented in Figure~\ref{fig:generalization}. 
Obtained results show that even the performance estimator makes some errors in terms of the criterion value, it can estimate the shape of the performance landscape with reasonable precision, based on the information about the model gained from the test run. This allows the proposed approach to tune the ESO in a near-optimal way even far beyond the scope of the training set. Even a predictor trained on a completely different system (NS vs. M1D) produces reasonable hints for observer gains tuning.

Analysis of the shapes of the ground truth performance landscapes for these two systems suggests that, while bandwidth parametrization allows for obtaining the whole spectrum of the performances for analyzed criteria, optimal solutions are achievable even when only two eigenvalues $\lambda_i$ are equal, and the third one may be different. Interestingly, the valley of the near-optimal performance lies on the line $\lambda_{1,2} + \lambda_3 = c$, where $c$ is some constant.

\begin{figure*}[h!]
    \centering
    \includegraphics[width=0.8\linewidth]{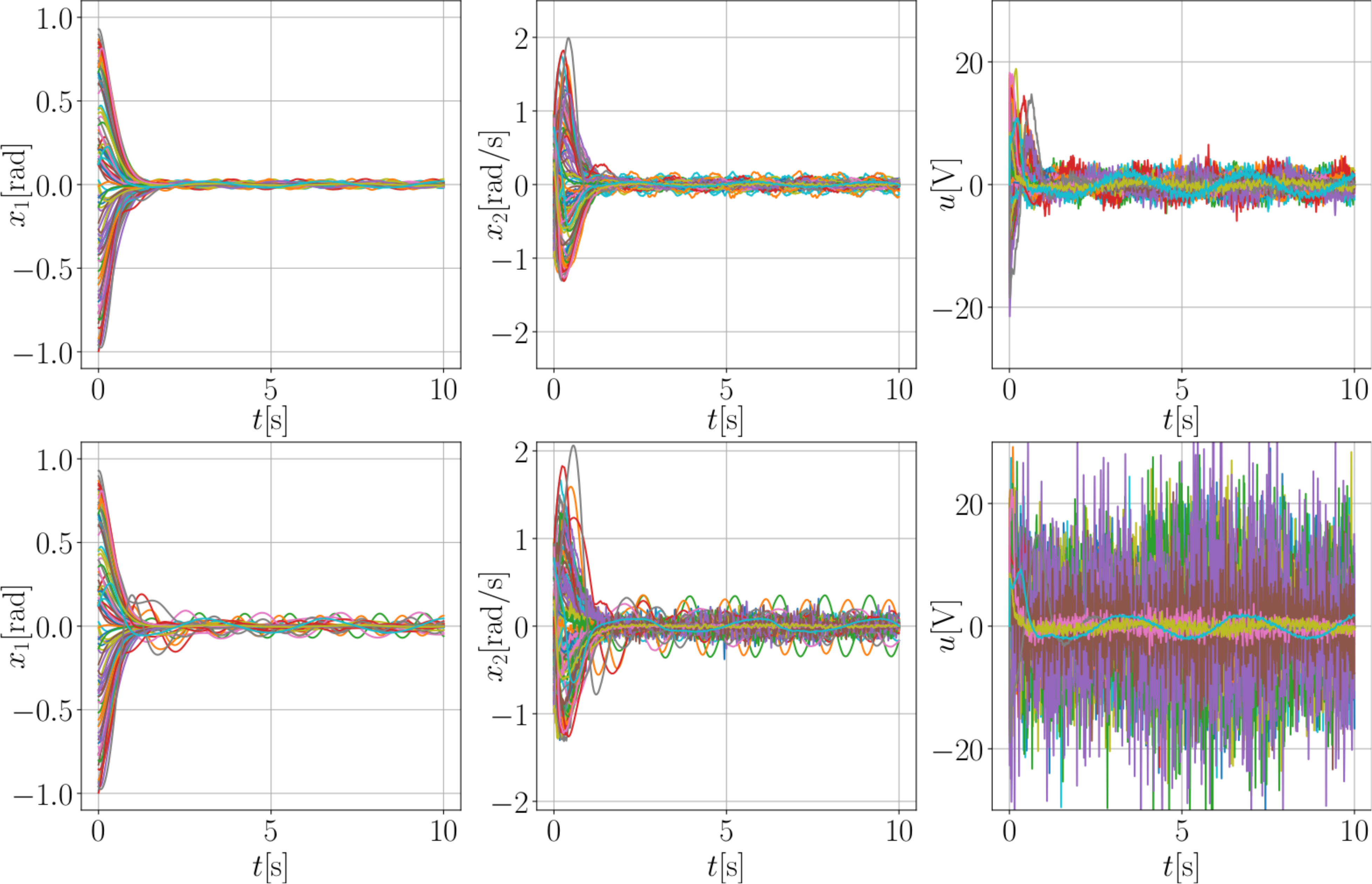}
    \caption{Transient states for a hundred of random NS objects. Top row contains transients for ADRC with ESO tuned with the proposed method for the performance criterion 0.989 IAE + 0.01 IAC + 0.001 IACD, while bottom with random $\omega_0$ value. Notice that, the neural network used to tune ESO gains was trained on M1D dataset.}
    \label{fig:generalization_monte_carlo}
\end{figure*}

A closer look at the performance of the ESO tuned with the use of the neural network trained on the different dataset is presented in Figure~\ref{fig:generalization_monte_carlo}. Here, we used a neural network trained on the M1D dataset to predict the observer gains for the set of 100 random NS systems. Resultant transients, compared with a random bandwidth parametrization, suggest that the proposed method can outperform the naive approach to ESO tuning, even when trained for a different system.

\section{Conclusions}
\label{sec:conclusions}

We presented a novel extended state observer tuning method for application in ADRC. The proposed solution uses a neural-network as a performance estimator of the closed loop system, which allows reasoning about the effects of the different choices of ESO gains. This method allows for multi-criteria ESO tuning, which takes into account not only the absolute error but also the control effort and the measurement noise suppression abilities.
The proposed tuning solution enables cheap checking of many possible parameter sets, without the need to repeatedly simulate or run the whole system, which saves time of the control engineer and avoids possibly costly tuning rehearsals on the real system. The important advantage of this solution is that it requires a single run of the closed-loop system to estimate its performance at any number of different ESO gains.

Our method was validated throughout several simulation experiments on two different, challenging control objects. Obtained results show that the proposed neural-network was able to estimate four integral performance metrics (IAE, IAC, IACD, IADEE), for previously unseen systems, with a mean absolute percentage error less than 4\%. This allowed us to show that our method can produce ESO gains that are near-optimal, in terms of the user-defined criterion. To demonstrate the quality of the tuning, we compared the transients of the systems tuned with the use of the proposed method, with the random choice of $\omega_0$ in the bandwidth parametrization, or several criteria combinations.

As the criterion is a user-defined combination of the selected integral control quality indicators, we analyzed their impact on the resultant eigenvalues of the state matrix in the closed-loop observation subsystem and the transients of the system tuned according to their combinations. The results showed that it is important to include the criteria related to the control effort in the tuning process, as it results in much more practically applicable control systems, and if one is not sure about the shape of the criterion, the Integral of Absolute Disturbance Estimation Error itself is a safe initial guess.

Moreover, as the proposed tuning method is data-driven, we showed its generalization abilities. 
Not only it works in a near-optimal fashion for the systems of the same structure with parameters out of the training data distribution, but even for systems of the very different structure it provides reasonably well tuning. Analysis of the performance landscapes for several different settings shows, that while the neural-network based performance estimator makes bigger errors for objects out of the training distributions, it still preserves the shape of the landscape, which is crucial for appropriate, near-optimal ESO tuning.

\vspace{-\baselineskip}
In future work, we would like to analyze the possibility to use the proposed system in an on-line fashion, to change the ESO gains every time the working conditions change. 
Moreover, the analysis of the performance landscapes suggests that, even though the bandwidth parametrization captures the near-optimal solution for the ESO tuning problem, one can propose a parametrization that will have better average performance and still lies near the optimal solution. This may be beneficial for the tuning ESO manually and seems like a promising future work direction.


\section{Acknowledgement}
The article was created thanks to participation in program PROM of the Polish National Agency for Academic Exchange. The program is co-financed from the European Social Fund within the Operational Program Knowledge Education Development, non-competitive project entitled “International scholarship exchange of PhD students and academic staff” executed under the Activity 3.3 specified in the application for funding of project No. POWR.03.03.00-00-PN13/18.

\appendix
\section{Proof of Theorem \ref{th:1}}
\label{app:proofTheorem1}

The analysis of the estimation error subsystem \eqref{eq:observationErrorDynamics} presented in this article has a local character for $\extendedStateVectorObservationError\in\ballDomain{\extendedStateVectorObservationErrorBallRadius}: 0<\extendedStateVectorObservationErrorBallRadius<\infty$, and is valid for perturbations $\totalDisturbanceDerivative\in\ballDomain{\totalDisturbanceDerivativeBallRadius}$ and $\measurementNoise\in\ballDomain{\measurementNoiseBallRadius}$, 
which boundedness is justified with Assumption \ref{ass:noiseBoundedness} and  Remark \ref{rem:totalDisturbance}. Let us introduce a positive definite function $\observationErrorLyapunovFunction\triangleq\extendedStateVectorObservationError^\top\lyapunovEquationSolution\extendedStateVectorObservationError:\ballDomain{\extendedStateVectorObservationErrorBallRadius}\rightarrow\nonNegativeRealNumbers$, limited by
\begin{align}
    \minEigenvalue{\lyapunovEquationSolution}\|\extendedStateVectorObservationError\|^2\leq\observationErrorLyapunovFunction(\extendedStateVectorObservationError)\leq\maxEigenvalue{\lyapunovEquationSolution}\|\extendedStateVectorObservationError\|^2.
    \label{eq:limitationLyapunovFunction}
\end{align}
Matrix $\lyapunovEquationSolution\succ0$ is a solution of the Lyapunov equation 
\begin{align}
    \observationErrorStateMatrix^\top\lyapunovEquationSolution + \lyapunovEquationSolution\observationErrorStateMatrix = -2\identityMatrix,
    \label{eq:lyapunovEquation}
\end{align}
and its positive-definiteness is guaranteed by $\observationErrorStateMatrix$ being Hurwitz, assured by the constrained values of the observer gain described in \eqref{eq:observerTuning1}-\eqref{eq:observerTuning3}. 
The derivative of $\observationErrorLyapunovFunction$, derived upon \eqref{eq:observationErrorDynamics} and \eqref{eq:lyapunovEquation}, can be expressed in a form
\begin{align}
    \observationErrorLyapunovFunctionDerivative &= \extendedStateVectorObservationError^\top(\observationErrorStateMatrix^\top\lyapunovEquationSolution+\lyapunovEquationSolution\observationErrorStateMatrix)\extendedStateVectorObservationError + 2\extendedStateVectorObservationError^\top\lyapunovEquationSolution\inputVector{3}\totalDisturbanceDerivative-2\extendedStateVectorObservationError^\top\lyapunovEquationSolution\observerGainVector\measurementNoise \nonumber \\ 
    &= -2\extendedStateVectorObservationError^\top\extendedStateVectorObservationError + 2\extendedStateVectorObservationError^\top\lyapunovEquationSolution\inputVector{3}\totalDisturbanceDerivative-2\extendedStateVectorObservationError^\top\lyapunovEquationSolution\observerGainVector\measurementNoise.
\end{align}
Next, let us derive 
\begin{align}
    \frac{d}{dt}\sqrt{\observationErrorLyapunovFunction} &= \frac{1}{2\sqrt{\observationErrorLyapunovFunction}}\observationErrorLyapunovFunctionDerivative \nonumber \\ 
    &= \frac{1}{\sqrt{\observationErrorLyapunovFunction}}\left[-\extendedStateVectorObservationError^\top\extendedStateVectorObservationError + \extendedStateVectorObservationError^\top\lyapunovEquationSolution\inputVector{3}\totalDisturbanceDerivative-\extendedStateVectorObservationError^\top\lyapunovEquationSolution\observerGainVector\measurementNoise\right] \nonumber \\ 
    &\leqtext{\eqref{eq:limitationLyapunovFunction}}-\frac{1}{\maxEigenvalue{\lyapunovEquationSolution}}\sqrt{\observationErrorLyapunovFunction}+\frac{\maxEigenvalue{\lyapunovEquationSolution}}{\minEigenvalue{\lyapunovEquationSolution}}\left[|\totalDisturbanceDerivative| + \|\observerGainVector\||\measurementNoise|\right].
    \label{eq:sqrtLyapunovFunctionDynamics}
\end{align}
Using the substitution 
\begin{align}
    \substitutedState(t) \triangleq  e^{\frac{1}{\maxEigenvalue{\lyapunovEquationSolution}} t}\sqrt{\observationErrorLyapunovFunction(t)},
    \label{eq:substitution}
\end{align}
within \eqref{eq:sqrtLyapunovFunctionDynamics}, we obtain
\begin{align}
    \frac{d}{dt}\left[e^{-\frac{1}{\maxEigenvalue{\lyapunovEquationSolution}} t}\substitutedState(t)\right] \leq & -\frac{1}{\maxEigenvalue{\lyapunovEquationSolution}} e^{-\frac{1}{\maxEigenvalue{\lyapunovEquationSolution}} t}\substitutedState \nonumber \\ &+\frac{\maxEigenvalue{\lyapunovEquationSolution}}{\minEigenvalue{\lyapunovEquationSolution}}\left[|\totalDisturbanceDerivative| + \|\observerGainVector\||\measurementNoise|\right],
\end{align}
implying
\begin{align}
    \substitutedStateDerivative(t) \leq e^{\frac{1}{\maxEigenvalue{\lyapunovEquationSolution}}t}\frac{\maxEigenvalue{\lyapunovEquationSolution}}{\minEigenvalue{\lyapunovEquationSolution}}\left[|\totalDisturbanceDerivative| + \|\observerGainVector\||\measurementNoise|\right].
    \label{eq:preint}
\end{align}
An upper-bound of state $\substitutedState(t)$ results from the integration of \eqref{eq:preint} and can be expressed as
\begin{align}
    \substitutedState(t) &\leq \substitutedState(0) + \frac{\maxEigenvalue{\lyapunovEquationSolution}}{\minEigenvalue{\lyapunovEquationSolution}}\int^t_0 e^{\frac{1}{\maxEigenvalue{\lyapunovEquationSolution}}\tau}\left[|\totalDisturbanceDerivative(\tau)| + \|\observerGainVector\||\measurementNoise(\tau)|\right]d\tau \nonumber \\ 
    &\leq \substitutedState(0) + \frac{\maxEigenvalue{\lyapunovEquationSolution}}{\minEigenvalue{\lyapunovEquationSolution}}\left[\sup_{t\geq0}|\totalDisturbanceDerivative(t)| + \|\observerGainVector\|\sup_{t\geq0}|\measurementNoise(t)|\right]\int^t_0 e^{\frac{1}{\maxEigenvalue{\lyapunovEquationSolution}}\tau}d\tau \nonumber \\ 
    &\leq \substitutedState(0) + \frac{\lambda_{\textrm{max}}^2({\lyapunovEquationSolution})}{\minEigenvalue{\lyapunovEquationSolution}}\left[\totalDisturbanceDerivativeBallRadius + \|\observerGainVector\|\measurementNoiseBallRadius\right]e^{\frac{1}{\maxEigenvalue{\lyapunovEquationSolution}}t},
\end{align}
where $\totalDisturbanceDerivativeBallRadius$ results from Remark \ref{rem:totalDisturbance}, and $\measurementNoiseBallRadius$ results from Assumption \ref{ass:noiseBoundedness}.
Now, going backwards with substitution \eqref{eq:substitution}, we obtain
\begin{align}
    e^{\frac{1}{\maxEigenvalue{\lyapunovEquationSolution}}t}\sqrt{\observationErrorLyapunovFunction(t)} &\leq \sqrt{\observationErrorLyapunovFunction(0)}+\frac{\lambda_{\textrm{max}}^2({\lyapunovEquationSolution})}{\minEigenvalue{\lyapunovEquationSolution}}\left[\totalDisturbanceDerivativeBallRadius + \|\observerGainVector\|\measurementNoiseBallRadius\right]e^{\frac{1}{\maxEigenvalue{\lyapunovEquationSolution}}t}, \label{eq:eqtrX}
\end{align}
and
\begin{align}
    \sqrt{\observationErrorLyapunovFunction(t)} &\leq e^{-\frac{1}{\maxEigenvalue{\lyapunovEquationSolution}}t}\sqrt{\observationErrorLyapunovFunction(0)}+\frac{\lambda_{\textrm{max}}^2({\lyapunovEquationSolution})}{\minEigenvalue{\lyapunovEquationSolution}}\left[\totalDisturbanceDerivativeBallRadius + \|\observerGainVector\|\measurementNoiseBallRadius\right]. \label{eq:eqtr}
\end{align}
Recalling \eqref{eq:limitationLyapunovFunction}, equation \eqref{eq:eqtr} can be rewritten as 
\begin{align}
    \|\extendedStateVectorObservationError(t)\|&\leq\frac{\maxEigenvalue{\lyapunovEquationSolution}}{\minEigenvalue{\lyapunovEquationSolution}}\|\extendedStateVectorObservationError(0)\|e^{-\frac{1}{\maxEigenvalue{\lyapunovEquationSolution}}t} + \frac{\lambda_{\textrm{max}}^2({\lyapunovEquationSolution})}{\lambda_{\textrm{min}}^2(\lyapunovEquationSolution)}\left[\totalDisturbanceDerivativeBallRadius + \|\observerGainVector\|\measurementNoiseBallRadius\right] \nonumber \\ 
    &:= c_1\|\extendedStateVectorObservationError(0)\|e^{-c_2t} + c_3\left[\totalDisturbanceDerivativeBallRadius + \|\observerGainVector\|\measurementNoiseBallRadius\right],
\end{align}
what completes the proof of Theorem \ref{th:1}.

\section{Proof of Theorem \ref{th:2}}
\label{app:proofTheorem2}

The analysis of the estimation error subsystem \eqref{eq:observationErrorDynamics} for the observer gain parametrization described by $\observerGainVector=[3\observerBandwidth \ 3\observerBandwidth^2 \ \observerBandwidth^3]^\top$ has a local character for $\extendedStateVectorObservationError\in\ballDomain{\extendedStateVectorObservationErrorBallRadius}: 0<\extendedStateVectorObservationErrorBallRadius<\infty$, and is valid for perturbations $\totalDisturbanceDerivative\in\ballDomain{\totalDisturbanceDerivativeBallRadius}$ and $\measurementNoise\in\ballDomain{\measurementNoiseBallRadius}$, which boundedness is justified with Assumption \ref{ass:noiseBoundedness} and  Remark \ref{rem:totalDisturbance}. In order to prove the desired features of closed-loop estimation system, we need to introduce a state transformation
\begin{align}
    \extendedStateVectorObservationError \triangleq \observationErrorEigenvalueMatrix\transformedExtendedStateVectorObservationError:\ballDomain{\transformedExtendedStateVectorObservationErrorBallRadius}\rightarrow\ballDomain{\extendedStateVectorObservationErrorBallRadius},
    \label{eq:observationErrorTransformation2}
\end{align}
where $\observationErrorEigenvalueMatrix\triangleq\textrm{diag}\{\observerBandwidth^{-2},\observerBandwidth^{-1},1\}$, and $\transformedExtendedStateVectorObservationErrorBallRadius\leq\max\{1,\observerBandwidth^{-2}\}\extendedStateVectorObservationErrorBallRadius$.
According to definition \eqref{eq:observationErrorTransformation2}, subsystem \eqref{eq:observationErrorDynamics} can be rewritten to a form
\begin{align}
    \transformedExtendedStateVectorObservationErrorDerivative &= \observationErrorEigenvalueMatrix^{-1}\observationErrorStateMatrix\observationErrorEigenvalueMatrix\transformedExtendedStateVectorObservationError + \observationErrorEigenvalueMatrix^{-1}\inputVector{3}\totalDisturbanceDerivative - \observationErrorEigenvalueMatrix^{-1}\observerGainVector\measurementNoise \nonumber \\ 
    &= \observerBandwidth\modifiedObservationErrorStateMatrix\transformedExtendedStateVectorObservationError + \inputVector{3}\totalDisturbanceDerivative - \observationErrorEigenvalueMatrix^{-1}\observerGainVector\measurementNoise,
    \label{eq:transformedDynamics2}
\end{align}
where
\begin{align}
    \modifiedObservationErrorStateMatrix = \begin{bmatrix} -3 & 1 & 0 \\ -3 & 0 & 1 \\ -1 & 0 & 0 \end{bmatrix}.
\end{align}
Now, let us introduce a positive-definite function $\transformedObservationErrorLyapunovFunction\triangleq\transformedExtendedStateVectorObservationError^\top\transformedLyapunovEquationSolution\transformedExtendedStateVectorObservationError:\ballDomain{\transformedExtendedStateVectorObservationErrorBallRadius}\rightarrow\nonNegativeRealNumbers$ limited by
\begin{align}
    \minEigenvalue{\transformedLyapunovEquationSolution}\|\transformedExtendedStateVectorObservationError\|^2\leq\transformedObservationErrorLyapunovFunction\leq\maxEigenvalue{\transformedLyapunovEquationSolution}\|\transformedExtendedStateVectorObservationError\|^2,
    \label{eq:lim2}
\end{align}
where $\transformedLyapunovEquationSolution\succ0$ is a solution of the Lyapunov equation
\begin{align}
    {\modifiedObservationErrorStateMatrix}^\top\transformedLyapunovEquationSolution+\transformedLyapunovEquationSolution\modifiedObservationErrorStateMatrix=-2\identityMatrix.
    \label{eq:lyapEq2}
\end{align}
The derivative of $\transformedObservationErrorLyapunovFunction$, derived upon \eqref{eq:transformedDynamics2}, can be expressed in a form
\begin{align}
    \transformedObservationErrorLyapunovFunctionDerivative &= \observerBandwidth\transformedExtendedStateVectorObservationError^\top({\modifiedObservationErrorStateMatrix}^\top\transformedLyapunovEquationSolution+\transformedLyapunovEquationSolution\modifiedObservationErrorStateMatrix)\transformedExtendedStateVectorObservationError + 2\transformedExtendedStateVectorObservationError^\top\transformedLyapunovEquationSolution\inputVector{3}\totalDisturbanceDerivative - 2\transformedExtendedStateVectorObservationError^\top\transformedLyapunovEquationSolution\observationErrorEigenvalueMatrix^{-1}\observerGainVector\measurementNoise \nonumber \\ 
    &\eqtext{\eqref{eq:lyapEq2}} -2\observerBandwidth\transformedExtendedStateVectorObservationError^\top\transformedExtendedStateVectorObservationError+ 2\transformedExtendedStateVectorObservationError^\top\transformedLyapunovEquationSolution\left[\inputVector{3}\totalDisturbanceDerivative - \observationErrorEigenvalueMatrix^{-1}\observerGainVector\measurementNoise\right].
\end{align}
Next, let us derive 
\begin{align}
    \frac{d}{dt}\sqrt{\transformedObservationErrorLyapunovFunction} &= \frac{1}{2\sqrt{\transformedObservationErrorLyapunovFunction}}\transformedObservationErrorLyapunovFunctionDerivative \nonumber \\ 
    &= \frac{1}{\sqrt{\transformedObservationErrorLyapunovFunction}}\left(-\observerBandwidth\transformedExtendedStateVectorObservationError^\top\transformedExtendedStateVectorObservationError+ \transformedExtendedStateVectorObservationError^\top\transformedLyapunovEquationSolution\left[\inputVector{3}\totalDisturbanceDerivative - \observationErrorEigenvalueMatrix^{-1}\observerGainVector\measurementNoise\right]\right) \nonumber \\ 
    &\leqtext{\eqref{eq:lim2}} -\frac{\observerBandwidth}{\maxEigenvalue{\transformedLyapunovEquationSolution}}\sqrt{\transformedObservationErrorLyapunovFunction}+\frac{\maxEigenvalue{\transformedLyapunovEquationSolution}}{\minEigenvalue{\lyapunovEquationSolution}}\left[|\totalDisturbanceDerivative| + 3\observerBandwidth^3|\measurementNoise|\right].
    \label{eq:sqrtLyapunovFunctionDynamics2}
\end{align}
Using the substitution 
\begin{align}
    \transformedSubstitutedState(t) \triangleq  e^{\frac{\observerBandwidth}{\maxEigenvalue{\transformedLyapunovEquationSolution}} t}\sqrt{\transformedObservationErrorLyapunovFunction(t)},
    \label{eq:substitution2}
\end{align}
within \eqref{eq:sqrtLyapunovFunctionDynamics2}, we obtain
\begin{align}
    \frac{d}{dt}\left[e^{-\frac{\observerBandwidth}{\maxEigenvalue{\transformedLyapunovEquationSolution}} t}\transformedSubstitutedState(t)\right] \leq & -\frac{\observerBandwidth}{\maxEigenvalue{\transformedLyapunovEquationSolution}} e^{-\frac{\observerBandwidth}{\maxEigenvalue{\transformedLyapunovEquationSolution}} t}\substitutedState(t) \nonumber\\ 
    &+\frac{\maxEigenvalue{\transformedLyapunovEquationSolution}}{\minEigenvalue{\lyapunovEquationSolution}}\left[|\totalDisturbanceDerivative| + 3\observerBandwidth^3|\measurementNoise|\right],
\end{align}
implying
\begin{align}
    \transformedSubstitutedStateDerivative(t) \leq e^{\frac{\observerBandwidth}{\maxEigenvalue{\transformedLyapunovEquationSolution}} t}\frac{\maxEigenvalue{\transformedLyapunovEquationSolution}}{\minEigenvalue{\lyapunovEquationSolution}}\left[|\totalDisturbanceDerivative| + 3\observerBandwidth^3|\measurementNoise|\right].
    \label{eq:preint2}
\end{align}
An upper-bound of state $\transformedSubstitutedState(t)$ results from the integration of \eqref{eq:preint2} and can be expressed as

\begin{align}
    \transformedSubstitutedState(t) &\leq \transformedSubstitutedState(0) + \frac{\maxEigenvalue{\transformedLyapunovEquationSolution}}{\minEigenvalue{\lyapunovEquationSolution}}\left[\sup_{t\geq0}|\totalDisturbanceDerivative(t)| + 3\observerBandwidth^3\sup_{t\geq0}|\measurementNoise(t)|\right]\int_0^t e^{\frac{\observerBandwidth}{\maxEigenvalue{\transformedLyapunovEquationSolution}} \tau}d\tau \nonumber \\ 
    &\leq \transformedSubstitutedState(0) + \frac{1}{\observerBandwidth}\frac{\lambda_{\textrm{max}}^2({\transformedLyapunovEquationSolution})}{\minEigenvalue{\lyapunovEquationSolution}}\left[\sup_{t\geq0}|\totalDisturbanceDerivative(t)| + 3\observerBandwidth^3\sup_{t\geq0}|\measurementNoise(t)|\right] e^{\frac{\observerBandwidth}{\maxEigenvalue{\transformedLyapunovEquationSolution}} t}.
\end{align}
Now, going backwards with substitution \eqref{eq:substitution2}, we obtain
\begin{align}
    \sqrt{\transformedObservationErrorLyapunovFunction(t)}\leq e^{-\frac{\observerBandwidth}{\maxEigenvalue{\transformedLyapunovEquationSolution}} t}\sqrt{\transformedObservationErrorLyapunovFunction(0)} + \frac{1}{\observerBandwidth}\frac{\lambda_{\textrm{max}}^2({\transformedLyapunovEquationSolution})}{\minEigenvalue{\lyapunovEquationSolution}}\left[\totalDisturbanceDerivativeBallRadius + 3\observerBandwidth^3\measurementNoiseBallRadius\right].
    \label{eq:eqtr2}
\end{align}
Recalling \eqref{eq:lim2}, equation \eqref{eq:eqtr2} can be rewritten as 
\begin{align}
    \|\transformedExtendedStateVectorObservationError(t)\|\leq e^{-\frac{\observerBandwidth}{\maxEigenvalue{\transformedLyapunovEquationSolution}} t}\frac{\maxEigenvalue{\transformedLyapunovEquationSolution}}{\minEigenvalue{\transformedLyapunovEquationSolution}}\|\transformedExtendedStateVectorObservationError(0)\| + \frac{1}{\observerBandwidth}\frac{\lambda_{\textrm{max}}^2({\transformedLyapunovEquationSolution})}{\lambda_{\textrm{min}}^2({\transformedLyapunovEquationSolution})}\left[\totalDisturbanceDerivativeBallRadius + 3\observerBandwidth^3\measurementNoiseBallRadius\right].
\end{align}
Finally, referring to transformation \eqref{eq:observationErrorTransformation2}, we may write an upper bound of the original observation error
\begin{align}
    \|\extendedStateVectorObservationError(t)\|&\leq\max\{\observerBandwidth^{-2},1\}\|\transformedExtendedStateVectorObservationError(t)\| \nonumber \\ 
    &\leq\max\{\observerBandwidth^{-2},\observerBandwidth^{2}\}e^{-c_4\observerBandwidth t}c_5\|\extendedStateVectorObservationError(0)\| \nonumber \\
    &+ \max\{\observerBandwidth^{-2},1\}\frac{1}{\observerBandwidth}c_6\left[\totalDisturbanceDerivativeBallRadius + 3\observerBandwidth^3\measurementNoiseBallRadius\right],
    \label{eq:proof2}
\end{align}
where $c_4=1/\maxEigenvalue{\transformedLyapunovEquationSolution}$, $c_5 = \maxEigenvalue{\transformedLyapunovEquationSolution}/\minEigenvalue{\transformedLyapunovEquationSolution}$, and $c_6=[\maxEigenvalue{\transformedLyapunovEquationSolution}/\minEigenvalue{\transformedLyapunovEquationSolution}]^2$. Result \eqref{eq:proof2} completes the proof of Theorem \ref{th:2}.

\section{Proof of Theorem \ref{th:3}}
\label{app:proofTheorem3}

The analysis of the closed-loop system \eqref{eq:closedloop} for the controller  \eqref{eq:genericController}, feedback function \eqref{eq:feedbackController}, and parametrization \eqref{eq:par0}-\eqref{eq:par1} has a local character for $\stateVector\in\ballDomain{\stateBallRadius}$ (see Assumption \ref{ass:1}). The perturbation of \eqref{eq:closedloop}, i.e. the estimation error vector $\extendedStateVectorObservationError$, is bounded upon Theorem \ref{th:1}. In order to prove desired behavior of $\stateVector$, we first need to introduce a transformation 
\begin{align}
    \stateVector \triangleq \stateVectorEigenvalueMatrix\transformedStateVector: \ballDomain{\transformedStateVectorBallRadius}\rightarrow \ballDomain{\stateBallRadius},
    \label{eq:stateTransformation}
\end{align}
where $\stateVectorEigenvalueMatrix\triangleq\textrm{diag}\{k^{-1},1\}$ and $\transformedStateVectorBallRadius\leq\max\{1,k^{-1}\}\stateBallRadius$. According to \eqref{eq:stateTransformation}, dynamics \eqref{eq:closedloop} can be rewritten to a form
\begin{align}
    \transformedStateVectorDerivative &= \stateVectorEigenvalueMatrix^{-1}\left(\stateMatrix{2}-\inputVector{2}\pmb{k}\right)\stateVectorEigenvalueMatrix\transformedStateVector + \stateVectorEigenvalueMatrix^{-1}\inputVector{2}[\pmb{k} \ 1]\extendedStateVectorObservationError \nonumber \\ 
    &= \controllerParameter\transformedStateStateMatrix\transformedStateVector+\inputVector{2}[\pmb{k} \ 1]\extendedStateVectorObservationError,
    \label{eq:transferedStateDynamics}
\end{align}
where
\begin{align}
    \transformedStateStateMatrix \triangleq \begin{bmatrix} 0 & 1 \\ -1 & -2\end{bmatrix}.
\end{align}
Now, let us introduce a positive-definite function $\stateVectorLyapunovFunction\triangleq\transformedStateVector^\top\stateVectorLyapunovEquationSolution\transformedStateVector:\ballDomain{\transformedStateVectorBallRadius}\rightarrow\nonNegativeRealNumbers$ limited by 
\begin{align}
\minEigenvalue{\stateVectorLyapunovEquationSolution}\|\transformedStateVector\|^2\leq\stateVectorLyapunovFunction\leq\maxEigenvalue{\stateVectorLyapunovEquationSolution}\|\transformedStateVector\|^2    
\label{eq:limits3}
\end{align}
where $\stateVectorLyapunovEquationSolution\succ0$ is a solution of the Lyapunov equation 
\begin{align}
    {\transformedStateStateMatrix}^\top\stateVectorLyapunovEquationSolution+\stateVectorLyapunovEquationSolution\transformedStateStateMatrix=-2\identityMatrix.
\end{align}
The derivative of $\stateVectorLyapunovFunction$, derived upon \eqref{eq:transferedStateDynamics}, can be expressed in a form
\begin{align}
    \stateVectorLyapunovFunctionDerivative &= \controllerParameter\transformedStateVector^\top({\transformedStateStateMatrix}^\top\stateVectorLyapunovEquationSolution+\stateVectorLyapunovEquationSolution\transformedStateStateMatrix)\transformedStateVector+2\transformedStateVector^\top\stateVectorLyapunovEquationSolution\inputVector{2}[\pmb{k} \ 1]\extendedStateVectorObservationError \nonumber \\ 
    &= -2\controllerParameter\transformedStateVector^\top\transformedStateVector+2\transformedStateVector^\top\stateVectorLyapunovEquationSolution\inputVector{2}[\pmb{k} \ 1]\extendedStateVectorObservationError.
\end{align}
Next, let us derive 
\begin{align}
    \frac{d}{dt}\sqrt{\stateVectorLyapunovFunction} &= \frac{1}{2\sqrt{\stateVectorLyapunovFunction}}\stateVectorLyapunovFunctionDerivative \nonumber \\ 
    &= \frac{1}{\sqrt{\stateVectorLyapunovFunction}}\left[-\controllerParameter\transformedStateVector^\top\transformedStateVector+\transformedStateVector^\top\stateVectorLyapunovEquationSolution\inputVector{2}[\pmb{k} \ 1]\extendedStateVectorObservationError\right] \nonumber \\
    &\leq -\controllerParameter\frac{1}{\maxEigenvalue{\stateVectorLyapunovEquationSolution}}\sqrt{\stateVectorLyapunovFunction}+\frac{\maxEigenvalue{\stateVectorLyapunovEquationSolution}}{\minEigenvalue{\stateVectorLyapunovEquationSolution}}\max\{\controllerParameter,1\}\|\extendedStateVectorObservationError\|.
    \label{eq:stateVectorLyapunovFunctionDerivative}
\end{align}
Using the substitution 
\begin{align}
    \substitutedStateStateVector(t) = e^{\frac{\controllerParameter}{\maxEigenvalue{\stateVectorLyapunovEquationSolution}}t}\sqrt{\stateVectorLyapunovFunction(t)}
    \label{eq:sub3}
\end{align}
within \eqref{eq:stateVectorLyapunovFunctionDerivative}, we obtain 
\begin{align}
    \frac{d}{dt}\left[e^{-\frac{\controllerParameter}{\maxEigenvalue{\stateVectorLyapunovEquationSolution}}t}\substitutedStateStateVector(t)\right] \leq & -\frac{\controllerParameter}{\maxEigenvalue{\stateVectorLyapunovEquationSolution}}e^{-\frac{\controllerParameter}{\maxEigenvalue{\stateVectorLyapunovEquationSolution}}t}\substitutedStateStateVector(t) \nonumber\\
    &+\frac{\maxEigenvalue{\stateVectorLyapunovEquationSolution}}{\minEigenvalue{\stateVectorLyapunovEquationSolution}}\max\{\controllerParameter,1\}\|\extendedStateVectorObservationError(t)\|,
\end{align}
implying
\begin{align}
    \substitutedStateStateVectorDerivative(t) \leq  e^{\frac{\controllerParameter}{\maxEigenvalue{\stateVectorLyapunovEquationSolution}}t}\frac{\maxEigenvalue{\stateVectorLyapunovEquationSolution}}{\minEigenvalue{\stateVectorLyapunovEquationSolution}}\max\{\controllerParameter,1\}\|\extendedStateVectorObservationError(t)\|.
    \label{eq:tointegrate}
\end{align}
An upper-bound of state $\substitutedStateStateVector(t)$ results from the integration of \eqref{eq:tointegrate} and can be expressed as 
\begin{align}
    \substitutedStateStateVector(t) &\leq \substitutedStateStateVector(0) + \frac{\maxEigenvalue{\stateVectorLyapunovEquationSolution}}{\minEigenvalue{\stateVectorLyapunovEquationSolution}}\max\{\controllerParameter,1\}\int_0^t\|\extendedStateVectorObservationError(\tau)\|e^{\frac{\controllerParameter}{\maxEigenvalue{\stateVectorLyapunovEquationSolution}}\tau}d\tau \nonumber \\ 
    &\leq \substitutedStateStateVector(0) + \frac{1}{\controllerParameter}\frac{\lambda_{\textrm{max}}^2({\stateVectorLyapunovEquationSolution})}{\minEigenvalue{\stateVectorLyapunovEquationSolution}}\max\{\controllerParameter,1\}\sup_{t\geq0}\|\extendedStateVectorObservationError(t)\|e^{\frac{\controllerParameter}{\maxEigenvalue{\stateVectorLyapunovEquationSolution}}t}.
\end{align}
Now, going backwards with substitution \eqref{eq:sub3}, we obtain
\begin{align}
    \sqrt{\stateVectorLyapunovFunction(t)}\leq e^{-\frac{\controllerParameter}{\maxEigenvalue{\stateVectorLyapunovEquationSolution}}t}\sqrt{\stateVectorLyapunovFunction(0)} + \frac{1}{\controllerParameter}\frac{\lambda_{\textrm{max}}^2({\stateVectorLyapunovEquationSolution})}{\minEigenvalue{\stateVectorLyapunovEquationSolution}}\max\{\controllerParameter,1\}\sup_{t\geq0}\|\extendedStateVectorObservationError(t)\|. 
    \label{eq:donthaveideaforanothername}
\end{align}
Recalling \eqref{eq:limits3}, equation \eqref{eq:donthaveideaforanothername} can be rewritten as
\begin{align}
    \|\transformedStateVector(t)\|\leq e^{-\frac{\controllerParameter}{\maxEigenvalue{\stateVectorLyapunovEquationSolution}}t}\frac{\maxEigenvalue{\stateVectorLyapunovEquationSolution}}{\minEigenvalue{\stateVectorLyapunovEquationSolution}}\|\transformedStateVector(0)\| + \frac{1}{\controllerParameter}\frac{\lambda_{\textrm{max}}^2({\stateVectorLyapunovEquationSolution})}{\lambda_{\textrm{min}}^2({\stateVectorLyapunovEquationSolution})}\max\{\controllerParameter,1\}\sup_{t\geq0}\|\extendedStateVectorObservationError(t)\|.
\end{align}
Finally, according to \eqref{eq:stateTransformation}, we may write an upper bound of the original state of the dynamic system
\begin{align}
    \|\stateVector(t)\|&\leq\max\{\controllerParameter^{-1},1\}\|\transformedStateVector(t)\| \nonumber \\ 
    &\leq \max\{\controllerParameter^{-1},\controllerParameter\}c_1e^{-c_2\controllerParameter t}\|\stateVector(0)\| + \frac{1}{\controllerParameter}c_3\max\{\controllerParameter^{-1},\controllerParameter\}\sup_{t\geq0}\|\extendedStateVectorObservationError(t)\|,
    \label{eq:proof3}
\end{align}
for $c_1 = \maxEigenvalue{\stateVectorLyapunovEquationSolution}/\minEigenvalue{\stateVectorLyapunovEquationSolution}$, $c_2=1/\maxEigenvalue{\stateVectorLyapunovEquationSolution}$, and $c_3=[\maxEigenvalue{\stateVectorLyapunovEquationSolution}/\minEigenvalue{\stateVectorLyapunovEquationSolution}]^2$. Result \eqref{eq:proof3} completes the proof of Theorem \ref{th:3}.


\section{Architecture of the neural network based performance estimator}
\label{sec:appendix_nn}

\begin{figure*}[th]
    \centering
    \includegraphics[width=0.85\linewidth]{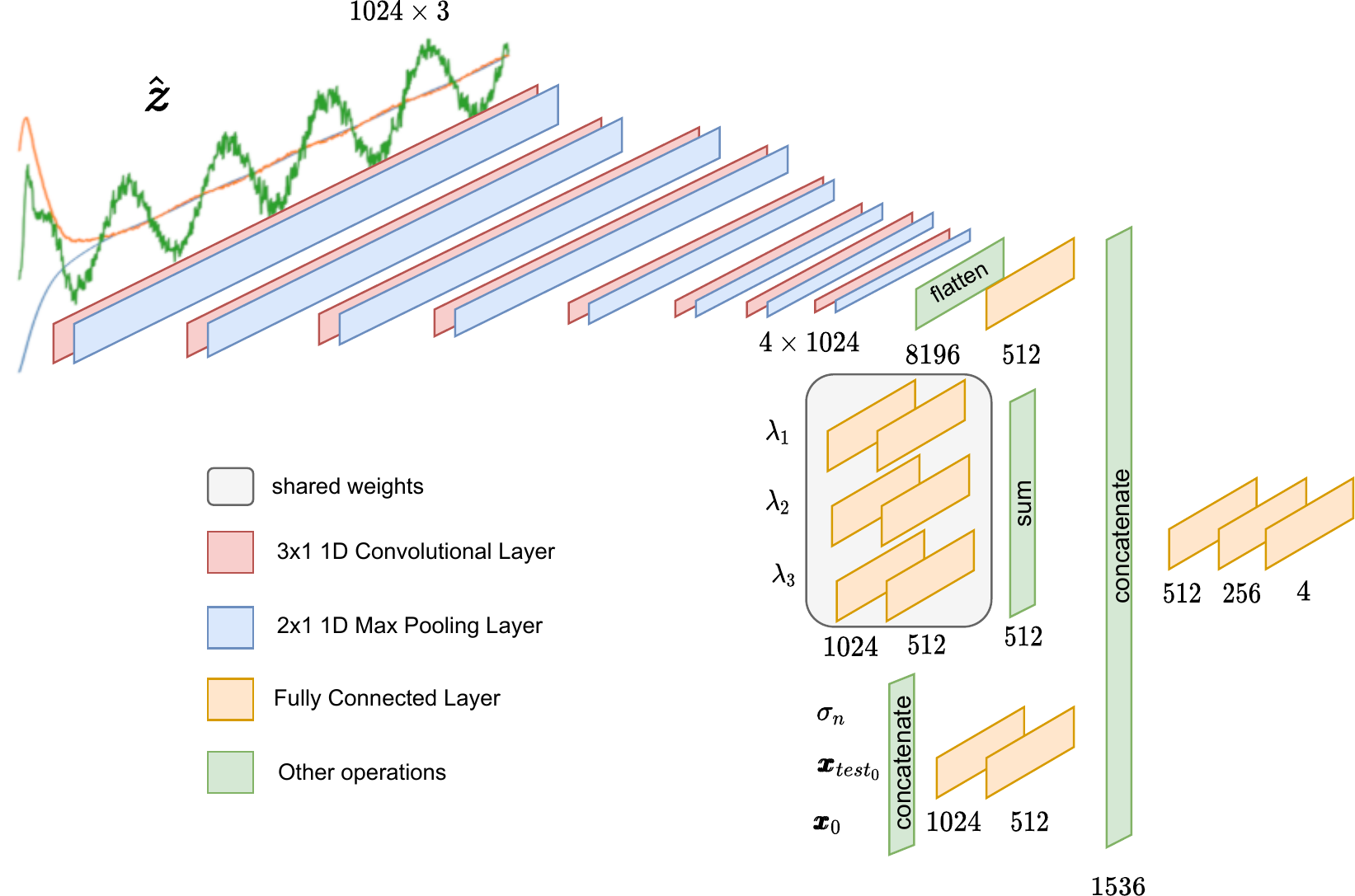}
    \caption{Detailed overview of the proposed performance estimator architecture.}
    \label{fig:detailed_architecture}
\end{figure*}

In Section \ref{sec:solution} we described the general scheme of the proposed neural network architecture, here we provide the details of this architecture. The overview of the architecture is presented in Figure~\ref{fig:detailed_architecture}. From the top we have a Basic experiment processing block, which is built from the 8 pairs of 3x1 1D Convolutional Layer, with Rectified Linear Unit (ReLU) activation \cite{relu}, and 2x1 1D Max Pooling Layers. The number of filters in consecutive convolutions doubles up, starting from 8 filters in the first layer up to the 1024 filters in the last convolutional layer. Finally, the output of convolutional blocks is flattened and passed to a single fully connected layer with 512 units.
Next, in the middle, there is a $\Lambda$ processing block presented, which consists of 3 processing streams, one for each $\lambda_i$, which shares weights for both fully connected layers. In the end, resultant feature vectors are summed up to achieve invariance to the permutation of the inputs. 
The last input processing block, which processes noise standard deviation as well as initial states for both experiments, performs firstly concatenation of inputs and then processes them with a two-layer fully connected neural network.
Outputs of those 3 blocks are concatenated together to create a common representation of the inputs. 
Finally, this representation is processed with 3 fully connected layers with 512, 256, and 4 units.
All fully connected layers, except the last one, are equipped with a ReLU activation. Output layer, in order to estimate the performance criteria of very different ranges, ends with a sigmoidal activation function, which returns values between 0 and 1, which are then scaled to fit ranges of the criteria.

\bibliography{bibliography}
\bibliographystyle{plain}

\end{document}

%% file: definitions.tex
\newcommand{\zeroMatrix}{\pmb{0}}
\newcommand{\identityMatrix}{\pmb{I}}
\newcommand{\realNumbers}{\mathbb{R}}
\newcommand{\complexNumbers}{\mathbb{C}}

\newcommand{\negativeRealNumbers}{\mathbb{R}_-}
\newcommand{\nonNegativeRealNumbers}{\mathbb{R}_{\geq0}}

\newcommand{\positiveIntegerNumbers}{\mathbb{Z}_+}

\newcommand{\minEigenvalue}[1]{\lambda_{\textrm{min}}({#1})}
\newcommand{\maxEigenvalue}[1]{\lambda_{\textrm{max}}({#1})}
\newcommand{\eigenvalue}[1]{\lambda_{#1}}
\newcommand{\eqtext}[1]{\overset{#1}{=}}
\newcommand{\leqtext}[1]{\overset{#1}{\leq}}

\newcommand{\realPart}[1]{\textrm{Re}\{#1\}}

\newcommand{\stateVector}{\pmb{x}}

\newcommand{\stateVectorEstimate}{\hat{\pmb{x}}}

\newcommand{\stateVectorElement}[1]{{x}_{#1}}

\newcommand{\stateVectorDerivative}{\dot{\pmb{x}}}
\newcommand{\systemOutput}{y}

\newcommand{\controlSignal}{u}

\newcommand{\extendedStateVector}{\pmb{z}}
\newcommand{\extendedStateVectorObservationError}{\tilde{\pmb{z}}}
\newcommand{\extendedStateVectorObservationErrorDerivative}{\dot{\tilde{\pmb{z}}}}
\newcommand{\extendedStateVectorDerivative}{\dot{\pmb{z}}}
\newcommand{\extendedStateVectorEstimate}{\hat{\pmb{z}}}
\newcommand{\extendedStateVectorEstimateDerivative}{\dot{\hat{\pmb{z}}}}
\newcommand{\extendedStateVectorElement}[1]{{z}_{#1}}

\newcommand{\observerGainVector}{\pmb{l}}
\newcommand{\observerGainVectorElement}[1]{{l}_{#1}}

\newcommand{\measurementNoise}{n}
\newcommand{\measurementNoiseStandardDeviation}{\sigma_n}
\newcommand{\totalDisturbance}{d}
\newcommand{\totalDisturbanceEstimate}{\hat{d}}
\newcommand{\totalDisturbanceDerivative}{\dot{d}}
\newcommand{\externalDisturbance}{d^{*}}
\newcommand{\externalDisturbanceDerivative}{\dot{d}^{*}}

\newcommand{\stateMatrix}[1]{\pmb{A}_{#1}}
\newcommand{\outputVector}[1]{\pmb{c}_{#1}}
\newcommand{\inputVector}[1]{\pmb{b}_{#1}}
\newcommand{\inputVectorExtendedState}[1]{\pmb{d}_{#1}}

\newcommand{\wenchaoDriftVector}{f}
\newcommand{\wenchaoInputVectorField}{g}
\newcommand{\wenchaoInputVectorFieldEstimate}{\hat{g}}
\newcommand{\wenchaoBenchmarkParameter}[1]{a_{#1}}
\newcommand{\pmxrBenchmarkParameter}[1]{b_{#1}}
\newcommand{\sawtooth}{\textrm{saw}}

\newcommand{\normalDistribution}{\mathcal{N}}

\newcommand{\ballDomain}[1]{\mathcal{B}_{#1}}
\newcommand{\stateBallRadius}{\bar{x}}
\newcommand{\measurementNoiseBallRadius}{\bar{n}}
\newcommand{\externalDisturbanceBallRadius}{\bar{d}^*}
\newcommand{\driftVectorBallRadius}{\bar{f}}
\newcommand{\totalDisturbanceBallRadius}{\bar{d}}
\newcommand{\totalDisturbanceDerivativeBallRadius}{\bar{D}}
\newcommand{\inputGainBallRadius}{\bar{g}}
\newcommand{\externalDisturbanceDerivativeBallRadius}{\bar{D}^*}
\newcommand{\continuousFunctionSet}[1]{\mathcal{C}^{#1}}

\newcommand{\extendedStateVectorObservationErrorBallRadius}{\bar{z}}
\newcommand{\transformedExtendedStateVectorObservationErrorBallRadius}{\bar{\zeta}}

\newcommand{\transformedExtendedStateVectorObservationError}{\pmb{\zeta}}
\newcommand{\transformedExtendedStateVectorObservationErrorDerivative}{\dot{\pmb{\zeta}}}

\newcommand{\observationErrorEigenvalueMatrix}{\pmb{\Lambda}_\zeta}
\newcommand{\stateVectorEigenvalueMatrix}{\pmb{\Lambda}_\epsilon}
\newcommand{\transformedStateVector}{\pmb{\epsilon}}
\newcommand{\transformedStateVectorDerivative}{\dot{\pmb{\epsilon}}}
\newcommand{\transformedStateVectorBallRadius}{\bar{\epsilon}}

\newcommand{\stateVectorEquilibrium}{\stateVector^*}
\newcommand{\feedbackController}{v}
\newcommand{\observationErrorStateMatrix}{\pmb{H}}
\newcommand{\modifiedObservationErrorStateMatrix}{\pmb{H}^*_\zeta}
\newcommand{\transformedStateStateMatrix}{\pmb{H}^*_\epsilon}

\newcommand{\observationErrorLyapunovFunction}{V_{\tilde{z}}}
\newcommand{\observationErrorLyapunovFunctionDerivative}{\dot{V}_{\tilde{z}}}

\newcommand{\stateVectorLyapunovFunction}{V_\epsilon}
\newcommand{\stateVectorLyapunovFunctionDerivative}{\dot{V}_\epsilon}

\newcommand{\transformedObservationErrorLyapunovFunction}{V_\zeta}
\newcommand{\transformedObservationErrorLyapunovFunctionDerivative}{\dot{V}_\zeta}

\newcommand{\substitutedState}{{\nu}_{\tilde{z}}}
\newcommand{\substitutedStateDerivative}{\dot{{\nu}}_{\tilde{z}}}
\newcommand{\substitutedStateStateVector}{{\nu}_\epsilon}
\newcommand{\substitutedStateStateVectorDerivative}{\dot{{\nu}}_\epsilon}
\newcommand{\transformedSubstitutedState}{{\nu}_\zeta}
\newcommand{\transformedSubstitutedStateDerivative}{\dot{{\nu}}_\zeta}

\newcommand{\lyapunovEquationSolution}{\pmb{P}_{\tilde{z}}}
\newcommand{\transformedLyapunovEquationSolution}{\pmb{P}_\zeta}
\newcommand{\stateVectorLyapunovEquationSolution}{\pmb{P}_\epsilon}
\newcommand{\observerBandwidth}{\omega_o}

\newcommand{\controllerGainElement}[1]{k_{#1}}

\newcommand{\controllerParameter}{k}
\newcommand{\costFunction}{J}
\newcommand{\costFunctionWeight}[1]{\alpha_{#1}}